\documentclass{article}
\usepackage{epsfig,graphicx}

\catcode`@=11
\def\secteqno{\@addtoreset{equation}{section}%
\def\theequation{\thesection.\arabic{equation}}}
\catcode`@=12

\def\Tr{{\rm ~Tr}\,}
\def\del{\partial}

\newcommand{\beq}{\begin{eqnarray}}
\newcommand{\eeq}{\end{eqnarray}}
\newcommand{\nn}{\nonumber}
\newcommand{\bit}{\begin{itemize}}
\newcommand{\eit}{\end{itemize}}
\newcommand{\ben}{\begin{enumerate}}
\newcommand{\een}{\end{enumerate}}
\newcommand{\bce}{\begin{center}}
\newcommand{\ece}{\end{center}}
\newcommand{\bfig}{\begin{figure}}
\newcommand{\efig}{\end{figure}}
\newcommand{\bfr}{\begin{flushright}}
\newcommand{\efr}{\end{flushright}}
\newcommand{\bfl}{\begin{flushleft}}
\newcommand{\efl}{\end{flushleft}}
\newcommand{\barr}{\begin{array}}
\newcommand{\earr}{\end{array}}
\newcommand{\abs}[1] { \left| #1 \right|}     
\newcommand{\kbra}[1] { \left< #1 \right>}     

\def\bm#1{\mbox{\boldmath${#1}$}}

\def\wb#1{\vbox{\ialign{##\crcr%
          \hskip 1.0pt\hrulefill\hskip 0.3pt%
          \crcr\noalign{\kern-1pt\vskip0.07cm\nointerlineskip}%
          $\hfil\displaystyle{#1}\hfil$\crcr}}}

\begin{document}

\begin{flushright}
NIIG-DP-03-1\\
June 10, 2003\\
hep-lat-0306003
\end{flushright}

\vskip .5in

\begin{center}
{\Large  Five-dimensional Lattice Gauge Theory as Multi-Layer World}\\
\vskip .75in

{\large  Michika Murata and Hiroto So}\\
\vskip .5in
{\it Department of Physics},
{\it Niigata University},
{\it Ikarashi 2-8050,
 Niigata 950-2181, Japan}
\end{center}

\vskip .5in

\begin{abstract}
A five-dimensional lattice space is decomposed into a number of four-dimensional  
lattices called  `layers'.  
We can interpret the five-dimensional gauge theory as 
four-dimensional gauge theories  on the multi-layer 
with interactions between  neighboring layers. 
In the theory, there exist two independent coupling constants:  
$\beta_4$ controls the dynamics inside a layer,  and $\beta_5$ does 
the strength of the inter-layer interaction.
We propose the new possibility to realize   the continuum limit of 
a five-dimensional  theory using 
 four-dimensional dynamics with large  $\beta_4$ and small $\beta_5$. 
Our result is also related to  the higher dimensional theory
 by deconstruction approach.
\end{abstract}

PACS~number(s):~11.10.Kk,~11.15.Ha,~11.25.Mj

\setcounter{footnote}{0}

\section{Introduction}    
Kaluza and Klein (KK)'s idea  was a nice unification scenario
that a four-dimensional gravity and a Maxwell theory are derived 
from  a five-dimensional gravity\cite{K21}.  
But their approach was done within a classical framework and 
it is not so easy to realize its quantum theory. 
The first problem is that 
five-dimensional theories are generally unrenormalizable 
even if they are renormalizable theories in four dimensions. 
The another point is that the property of phase transitions in five-dimensional 
lattice theories is exactly same as  that of mean-field theories. 
The latter implies that it is generally difficult to take the continuum limit 
in the five-dimensional theory using the second order phase transition 
and the associated fixed point (FP). 
Actually, some searches for both properties 
in a five-dimensional gauge theory have been tried
\cite{L-P-S-87,K-N-O92,E-K-M00,E-F-K02}
 and it is still far that one  realizes {\it a continuum limit} 
from  five-dimensional  statistical mechanics. 

Although most of possibilities in  
the construction of five-dimensional theories are shut,   
there is a path to the construction, that is to use well-known properties of 
a four-dimensional  statistical mechanics. 
Since we know  the continuum limit of a four-dimensional gauge theory, 
the limiting operation keeping fixed physical quantities  can be considered in the meaning 
of a four-dimensional theory. 
Recently, a new approach for the KK's scenario called deconstruction has appeared 
using four-dimensional asymptotically free gauge theories 
in the continuum\cite{A-C-G01,C-H-P-W01}. 
The approach realizes KK-like modes 
although they are completely four-dimensional theories. 
It is an interesting idea but can not give any  answer for difficulties 
 of  higher dimensional theories because the idea is based on the perturbative 
picture of four-dimensional theories and  has nothing to any clue about  
properties of the statistical mechanics. 

On the other hand, Fu and Nielsen suggested  to reduce 
the extra-dimensional space  by  the dynamics of 
a higher-dimensional lattice gauge theory\cite{F-N84}. 
In the idea, they considered a decomposition of a five-dimensional lattice space 
into an ensemble of four-dimensional lattice spaces (multi-layer). 
Unfortunately, they studied about compact $U(1)$  gauge theories  
which have  no  continuum limit 
in four dimensions\cite{H-KA-N-95,D-F-KA-K-N-01,D-F-K-K-01,D-F-N-02}. 
In this paper, we formulate a five-dimensional lattice $SU(2)$ gauge  theory as 
a block of four-dimensional theories with an extra interaction like as Fu and Nielsen. 
The phase structure of the theory is clarified with Monte Carlo simulation and 
we discuss about the continuum limit as a four-dimensional multi-layer system 
(multi-layer world).
Finally, we suggest how to construct a five-dimensional theory by explicit limiting
operations.
A limiting curve corresponding to five-dimensional scale shall becomes to an envelope. 

The outline of our paper is as  follows:
our model setting and the notation are explained in the next section. 
In sec.3, our model is investigated by a numerical method. 
In sec.4, we discuss physical properties of our model such as   
gauge squared mass matrix,  the four-dimensional continuum limit 
and  suggestion of a new idea for defining a five-dimensional theory.  
Sec.5 is devoted to  summary and discussions. 

\section{Our Model} 

In this section, we shall set our model in addition to explanation for 
a layer structure in the five-dimensional lattice. 
 Link variables along a fifth direction are  interpreted as 
scalar fields in  the model. 
Naively, this has a picture of a four-dimensional theory  
in the continuum limit. Afterwards, measured order parameters in 
our numerical simulation are listed and  an expected phase diagram  is sketched. 

\begin{figure}[p]
\begin{center}
$\hspace{2.1cm}{\cal L}_1 \hspace{1cm}{\cal L}_2 \hspace{1.3cm}{\cal L}_{n_5}\hspace{1.5cm}{\cal L}_{N_5} \hspace{1cm}{\cal L}_1 $
\end{center}
\vspace{-2.25cm}
\begin{center}
\includegraphics[scale=0.6]{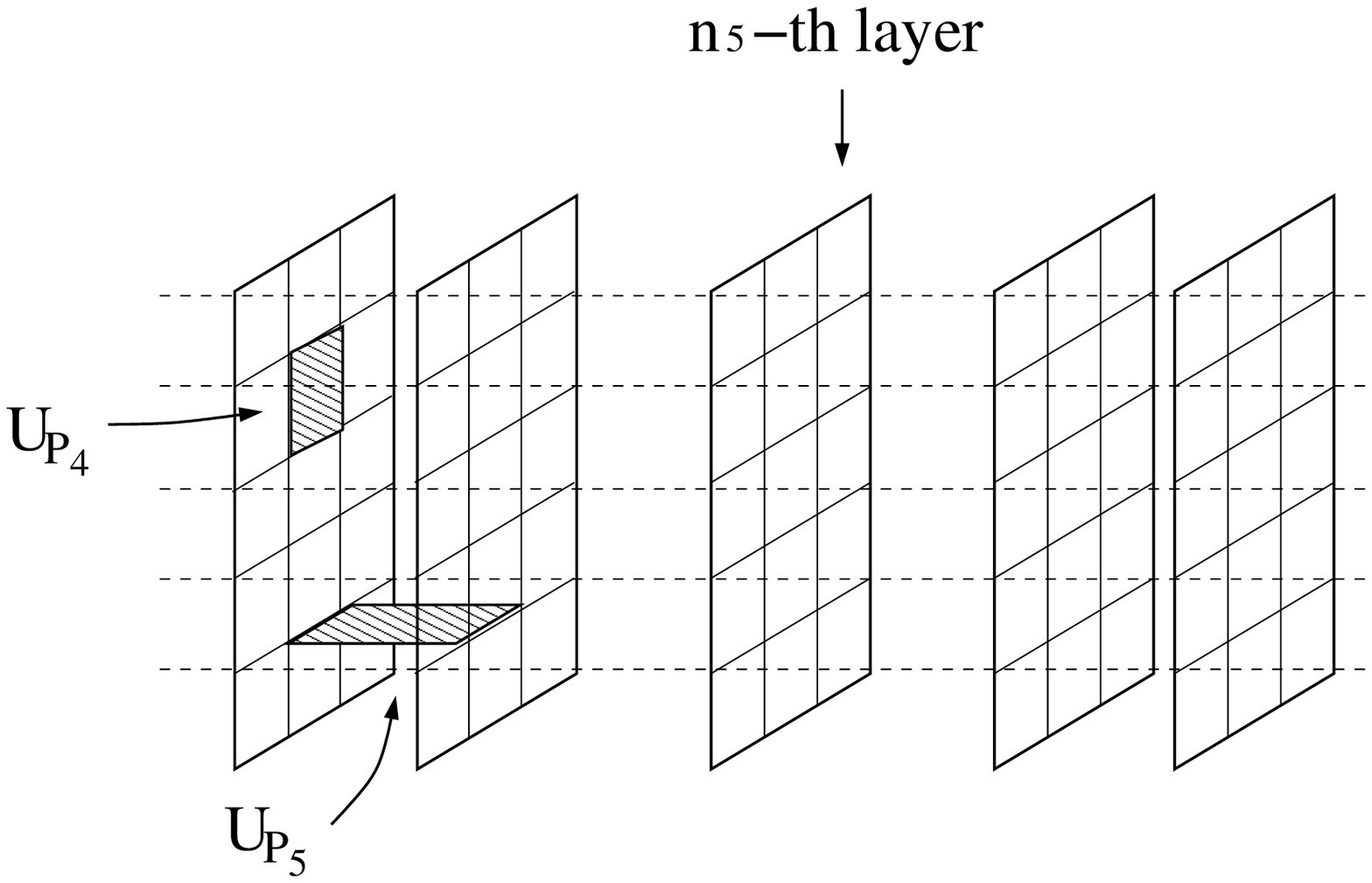}
\vspace{5cm}
\caption{Layer structure of five-dimensional lattice.}
\label{figlayer}
\end{center}
\end{figure}

\hspace{-0.5cm}{\bf 1.~Layer structure and Wilson action}
We consider the five-dimensional pure $SU(2)$ Wilson action on a $N_4^4 \times N_5$
 lattice with a periodic boundary condition:
\beq
S_{\rm lat}
&=&\beta_{4}\sum_{P_4}\left[1-\frac{1}{2}\Tr U_{P_4}\right]
  +\beta_{5}\sum_{P_5}\left[1-\frac{1}{2}\Tr U_{P_5}\right]\ ,
\label{lact}
\eeq
\beq
\Tr U_{P_4}
&\equiv& \Tr U_{n,\mu}^{n_5}\ U_{n+\hat{\mu},\nu}^{n_5}
      \ U_{n+\hat{\nu},\mu}^{\dagger n_5}\ U_{n,\nu}^{\dagger n_5}\ ,
\\[4pt]
\Tr U_{P_5}
&\equiv& \Tr U_{n,\mu}^{n_5}\ U_{n+\hat{\mu},5}^{n_5}
      \ U_{n,\mu}^{\dagger n_5+1}\ U_{n,5}^{\dagger n_5}\ ,
\eeq
where the sum is defined as
$\sum_{P_4} \equiv \sum_{n,n_5}\ \sum_{1\leq \mu < \nu \leq 4}\ ,\ 
\sum_{P_5} \equiv \sum_{n,n_5}\ \sum_{1\leq \mu \leq 4}$ . 
According to Fu and Nielsen\cite{F-N84}, we call the four-dimensional
 subspace a `layer',
 because the individual four-dimensional subspace
 must be independent of each other if $\beta_5$ is equal to zero.
Every layer was aligned along the fifth direction $-\!\!\!-$ see 
Fig.\ref{figlayer}. 
An $n_5$-th layer is noted ${\cal L}_{n_5}$.
$U_{n,\mu}^{n_5}$ is a link variable inside ${\cal L}_{n_5}$ 
 and  $U_{n,5}^{n_5}$ connects between ${\cal L}_{n_5}$ and ${\cal L}_{n_5+1}$.
These link variables behave as $\underline{2}$ of $SU(2)$.
Then the first term of the action (\ref{lact}) represents interactions inside layers
controlled by $\beta_4$, 
 and the second term represents interactions between different layers (inter-layer) 
controlled by $\beta_5$\footnote{
A similar model with the adjoint Higgs field 
is investigated\cite{D-F-K-02},  where the field plays  an  essential role 
in the existence of a  `layered' phase. Although we does not assume 
the phase in this work, it seems an interesting subject. 
}.
Note that for convenience two coupling constants,
\beq
\beta \equiv \sqrt{\beta_4 \beta_5}\ ,\ 
\gamma \equiv \sqrt{\frac{\beta_5}{\beta_4}} \ ,
\label{par}
\eeq
are also used.


\hspace{-0.5cm}{\bf 2.~Measuring quantities}
To understand the dynamics of (\ref{lact}), the following quantities are measured. 
\bit
\item
Expectation values of the plaquette $U_{P_4}$ inside a layer 
and $U_{P_5}$ between neighboring layers.
\item
Creutz ratio of loops inside a layer,
\beq
\sigma_4(l_{\mu},l_{\nu}) \equiv -{\rm ln} 
   \left\{ \frac{W(l_{\mu},l_{\nu})W(l_{\mu}-1,l_{\nu}-1)}
{W(l_{\mu},l_{\nu}-1)W(l_{\mu}-1,l_{\nu})}\right\}\ ,
\eeq
where $W(l_{\mu},l_{\nu})$ implies an expectation value of Wilson loop 
with $l_{\mu} \times l_{\nu}$ size on the same layer.
\item
Polyakov loop along fifth direction,
\beq
L_{5,n} \equiv \frac{1}{2}{\rm Tr }\prod_{n_5} U_{n,5}^{(n_5)}\ .
\label{Poly}
\eeq
\item
Inverse of correlation length for Polyakov loops,
\beq
\kbra{L_{5,n}L_{5,m}}-\kbra{L_{5,n}}^2 
 &\sim& \cosh(-H\abs{n-m})\ .
\label{Kdef}
\eeq
\eit
$U_{P_4}$ and $U_{P_5}$ correspond to parts of the internal energy.
$\sigma_4(l_{\mu},l_{\nu})$ is an order parameter of four-dimensional
 confinement inside a layer
and $\kbra{L_{5,n}}$ is that of $Z_2$ center symmetry
\footnote{
For the $N_5=1$ case, a singlet part of $U_{n,5}$ is important for
 the phase structure. For the realization of a non-trivial transformation for 
 the singlet part only, we add a special gauge transformation to the center symmetry.
}.

\hspace{-0.5cm}{\bf 3.~Intuitive understanding of dynamics}
Depending on the magnitude of $\beta_4, \beta_5$, 
the dynamics (\ref{lact}) has various feature. 
In the case that $\gamma$ is larger than unity, inter-layer interactions are 
strong compared to those inside layers.
In this region the influence among the layer can not be ignored because a layer is
 strongly bound to its neighboring layers.
It seems difficult to control the system in taking the continuum limit 
 because there remain the strong extra-interactions caused by the fifth direction 
 in the coupling region.
As $\gamma$ is smaller than unity, interactions of inter-layer are weaker than
 those of inside layers.
Then layers are loosely bound to each other,
 and in this region the dynamics is
 similar to a four-dimensional gauge theory with extra fields.
We call this picture a `multi-layer world'.

\begin{figure}[p]
\begin{center}
\includegraphics[scale=0.72]{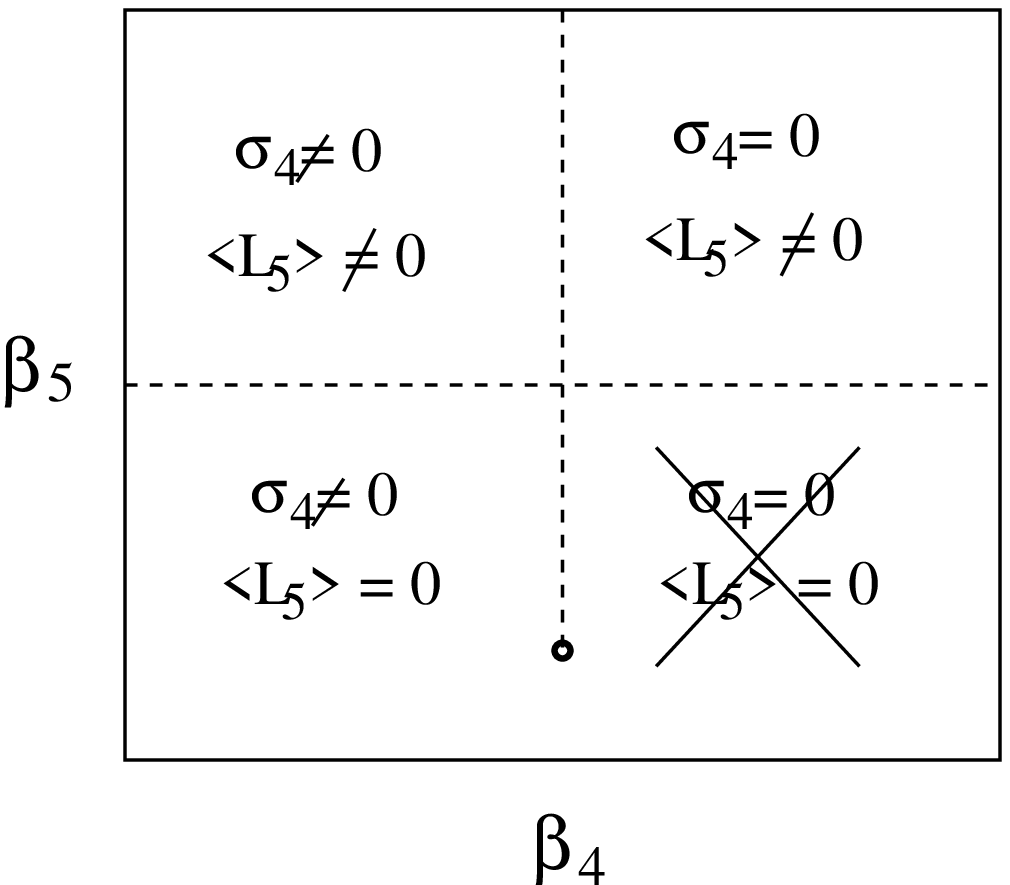}
\vspace{5cm}
\caption{Intuitive understanding of phase diagram in (2.1).}
\label{ponchi}
\end{center}
\end{figure}

From two order parameters $\sigma_4$ and $\kbra{L_5}$ we can imagine four phases. 
Since our model is reduced to a four-dimensional gauge theory at $\beta_5=0$,
 we know that the phase of $\sigma_4=0$ and $\kbra{L_5}=0$ is not realized 
$-\!\!\!-$ see Fig.\ref{ponchi}.


\hspace{-0.5cm}{\bf 4.~Naive continuum limit as four-dimensional gauge theory}
We consider the naive continuum limit of our model for a particular case which
 $U_{n,5}^{n_5}$ is interpreted as a fundamental `scalar' field $V^{n5}_{n}$ 
 on the ${\cal L}_{n_5}$.
In the interpretation, the fifth direction is simply an internal degree of freedom. 
We should consider limit $a_4 \rightarrow 0$ as the naive continuum limit
 because there is only lattice spacing $a_4$ inside layers. 
In the first term of the action (\ref{lact}) $\Sigma_{n_5}$ corresponds to a trace 
operation in the internal space and $\Sigma_n$ dose to the summation of 
four-dimensional space. 
Then the relation between the inside layer coupling $\beta_4$ and a four-dimensional 
gauge coupling  $g_4$ is determined as
\beq
\beta_{4} = \frac{4}{g_4^2}\ .
\eeq 
On the other hand, the second term of the action (\ref{lact})
means interactions between gauge field and `scalar' field written as
\beq
U_{P_5}
&=&  U_{n,\mu}^{n_5}\ V_{n+\hat{\mu}}^{n_5}
      \ U_{n,\mu}^{\dagger n_5+1}\ V_{n}^{\dagger n_5}\ .
\eeq
The $\beta_5$ is the inter-layer coupling which controls dynamics of `scalar' fields,
 and it is determined from the renormarization factor in the continuum limit.\\

\section{Numerical Analysis $-\!\!\!\!-\!\!\!\!-$ Phase Structures on Various $\bm{N_5}$ $-\!\!\!\!-\!\!\!\!-$}

In this section, we investigate the phase structure of
 our model by using Monte Carlo simulation.
We simulate for Wilson action (\ref{lact}) using $N_4^4
 \times N_5$ lattice ($N_4=8,10,12$, $N_5=1,2,3,5,6,8$) by heat-bath algorithm.
We measure  expectation values of plaquettes $\kbra{U_{P_4}}$, $\kbra{U_{P_5}}$,
 specific heats
 $C_4\equiv \left<U_{\rm P_4}^2\right>-\left<U_{\rm P_4}\right>^2$,
 $C_5\equiv \left<U_{\rm P_5}^2\right>-\left<U_{\rm P_5}\right>^2$,
 Creutz ratio and Polyakov loop with its fluctuation 
 $\chi_5 \equiv \left<L_5^2\right>-\left<L_5 \right>^2$.
The statistical errors of these measurements are estimated by the jackknife method.
In the following figures, open and filled symbols are noted as results of
 ordered and disordered start respectively.
We describe $N_5=1$ and $N_5 >1 $ cases separately.

\subsection{Case of $\bm{N_5=1}$}

\begin{figure}[p]
\begin{center}
\includegraphics[scale=0.5]{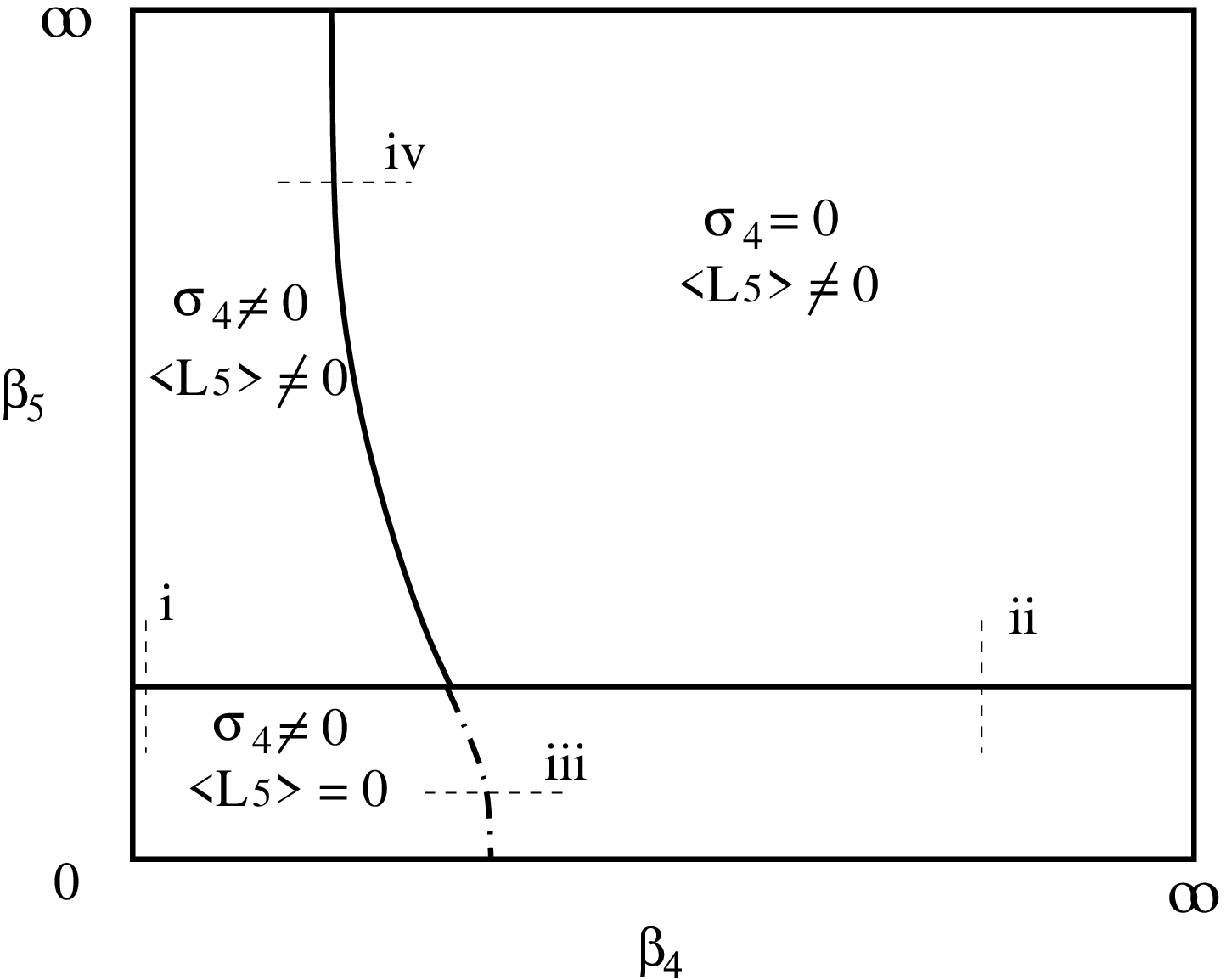}
\vspace{5cm}
\caption{
The phase diagram for $N_5=1$.
Solid lines imply phase boundaries and a dash-dotted line is originated in the crossover 
of a four-dimensional gauge theory.
Dashed lines correspond to measured directions. 
}
\label{phdiagN5=1}
\end{center}
\end{figure}

Fig.\ref{phdiagN5=1} represents the phase diagram at $N_5=1$ and plots measured 
lines where 
 (i) $\beta_4=0$, (ii) $\gamma=1/\sqrt{12}\sim 0.29$,
 (iii) $\beta_4=10.0$ and (iv) $\beta_5=10.0$.   
Our calculations support the expected phase diagram (Fig.\ref{ponchi}),
and their detailed discussions are summarized as follows:\\

\begin{figure}[p]
\includegraphics[scale=0.4]{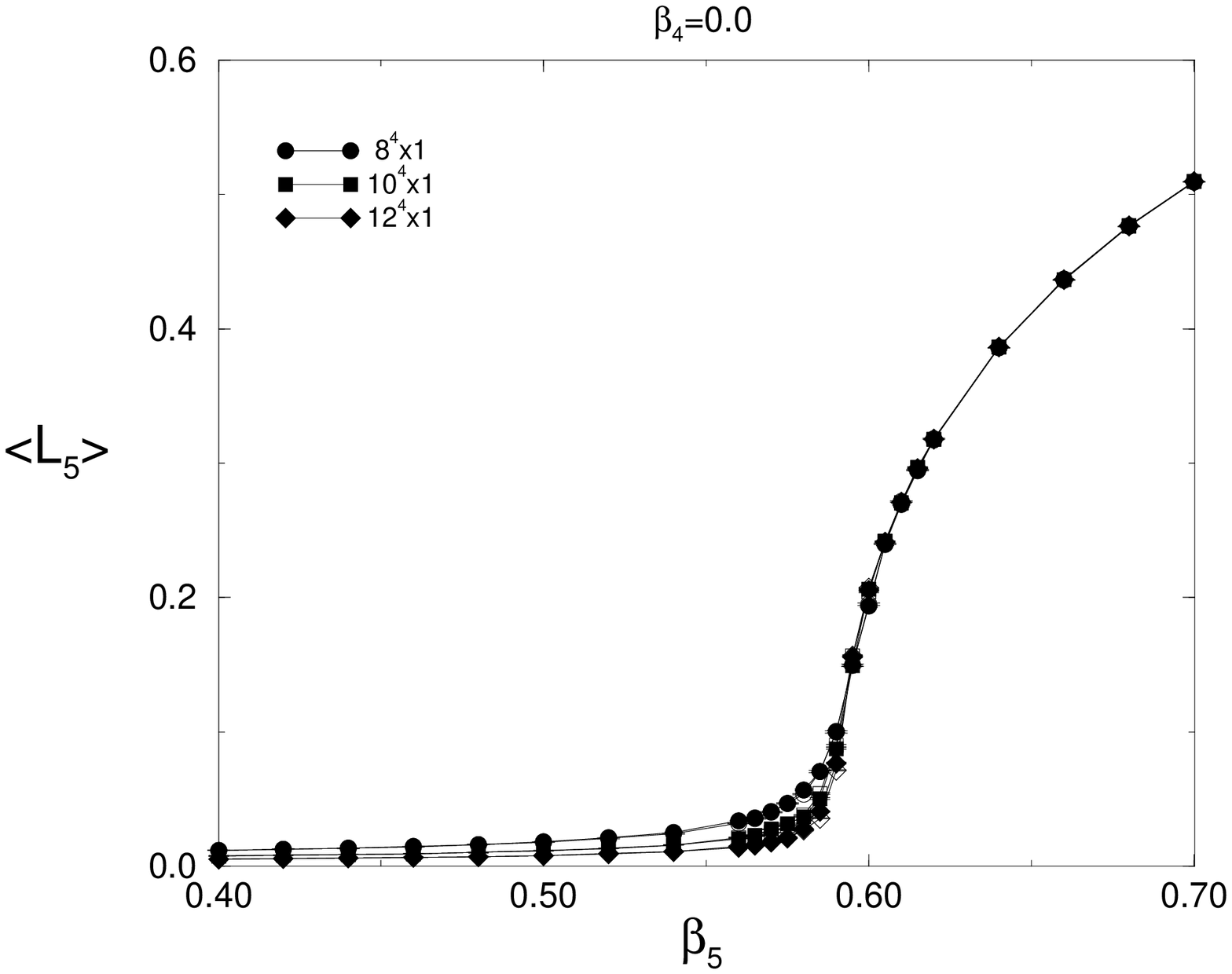}
\includegraphics[scale=0.4]{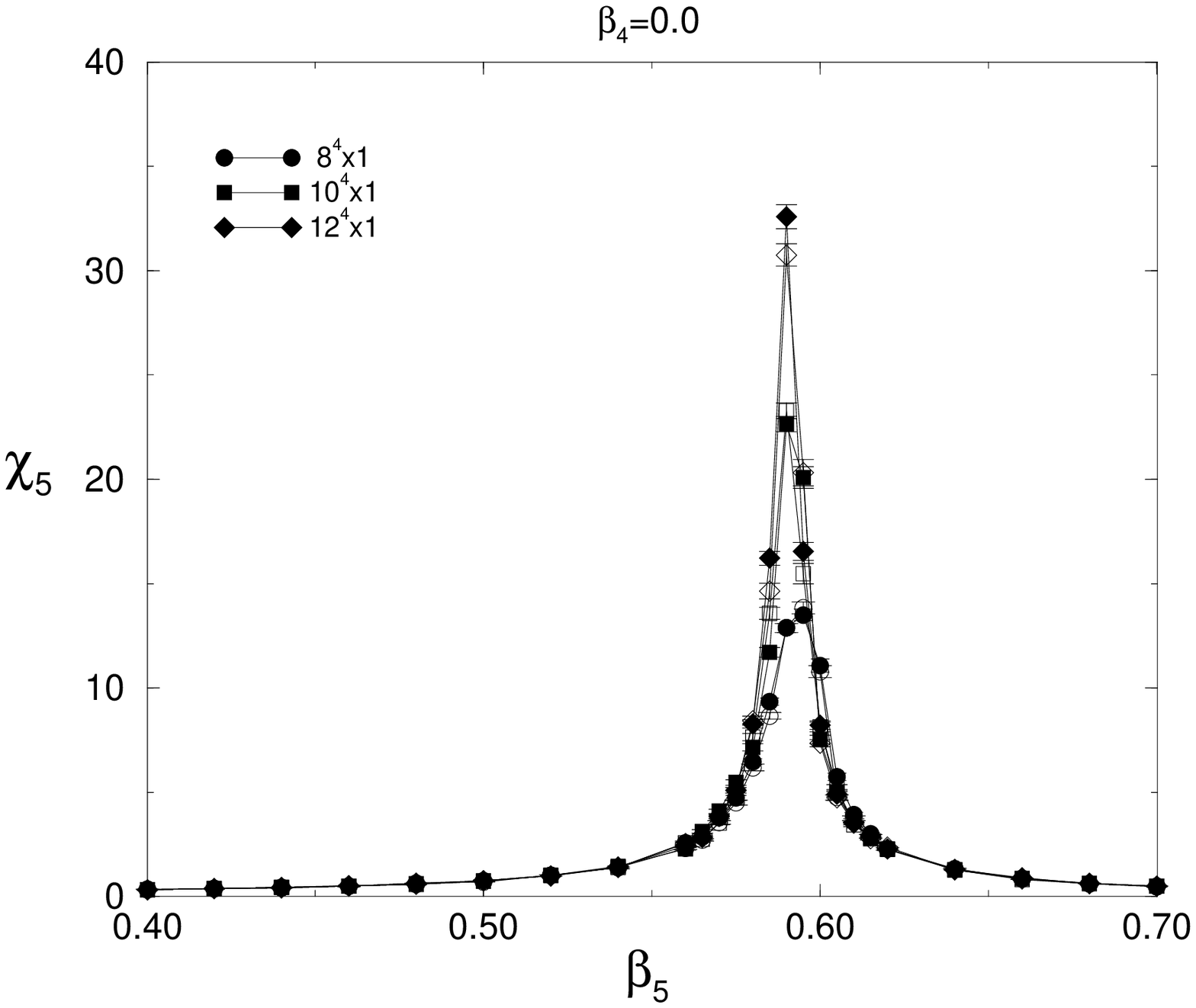}
\vspace{5cm}
\caption{$\kbra{L_5}$ and $\chi_5$ at $\beta_4=0$.}
\label{fign5=1b4-0}
\end{figure}

(i) ${\it \beta_4=0}$: 
Fig.\ref{fign5=1b4-0} shows results of $\kbra{L_5}$
 and $\chi_5$. 
Those quantities imply that the confinement-deconfinement phase transition occurs 
at $\beta_{5c} \sim 0.6$.
Especially in $\chi_5$, we can see the volume effect which is a characteristic signal
 of the phase transition.

\begin{figure}[p]
\includegraphics[scale=0.4]{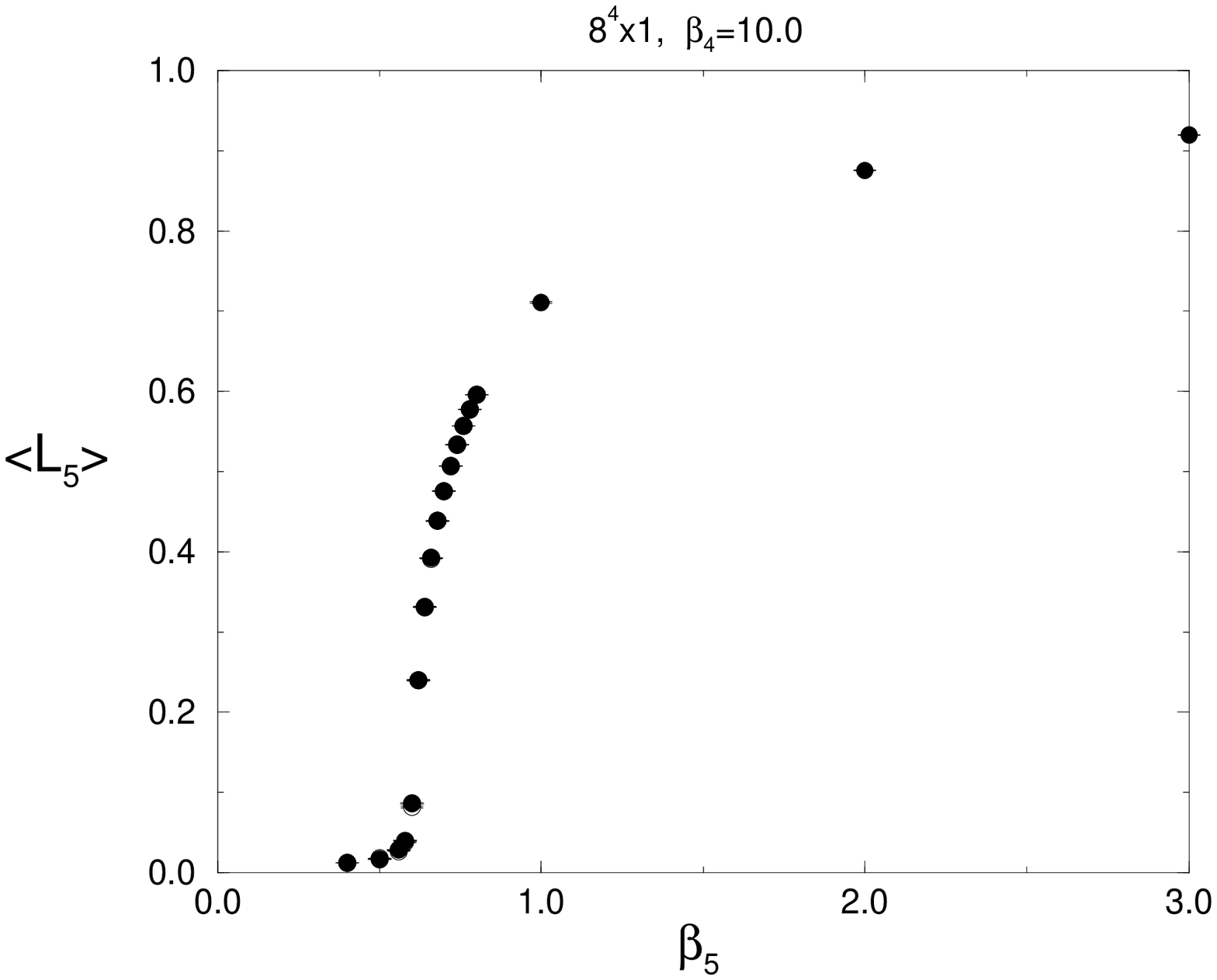}
\includegraphics[scale=0.4]{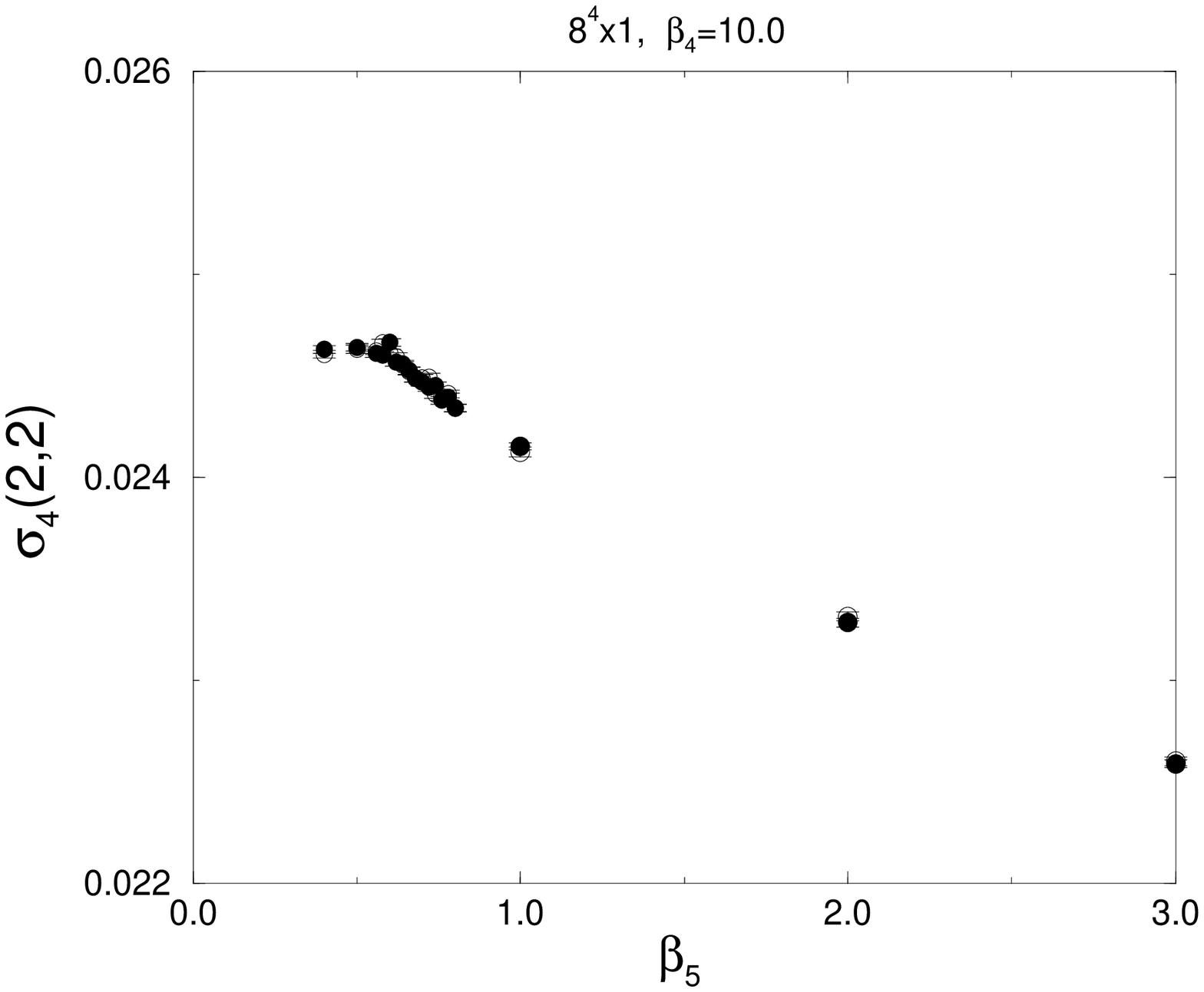}
\vspace{5cm}
\caption{$\kbra{L_5}$ and $\sigma_4(2,2)$ at $\beta_4=10.0$. }
\label{fign5=1b4-10}
\end{figure}

\vspace{1mm}
(ii) ${\it \beta_4=10.0}$: 
From $\kbra{L_5}$, the similar signal of the transition can be seen at $\beta_5 \sim 0.6$.
In addition, $\sigma_4$ also bends sharply at $\beta_5 \sim 0.6$
 $-\!\!\!-$ see Fig.\ref{fign5=1b4-10}.

\begin{figure}[p]
\begin{center}
\includegraphics[scale=0.4]{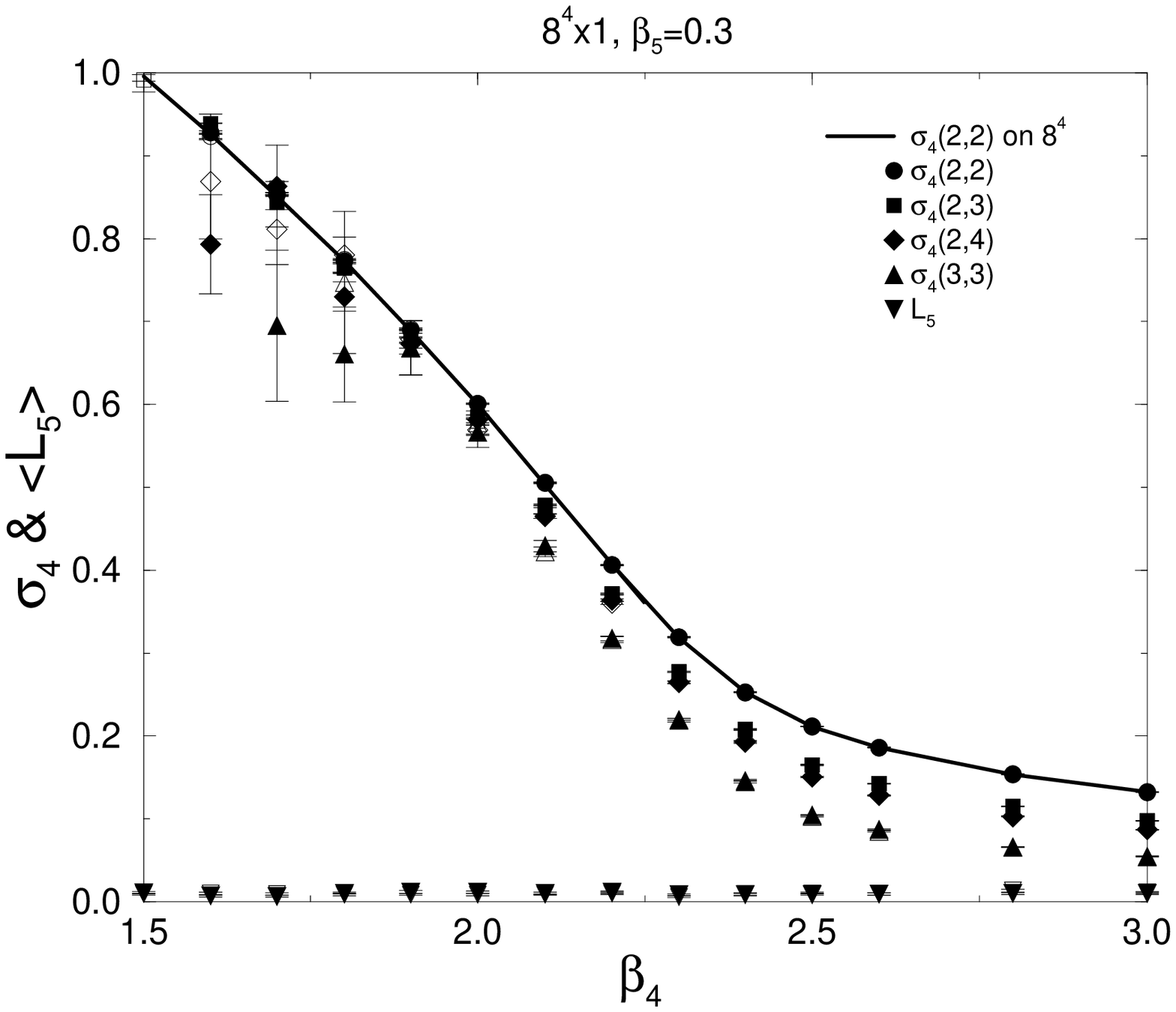}
\vspace{5cm}
\caption{$\sigma_4$ and $\kbra{L_5}$ at $\beta_5=0.3$.}
\label{fign5=1b5-03}
\end{center}
\end{figure}

\vspace{1mm}
(iii) ${\it \beta_5 =0.3}$: 
In Fig.\ref{fign5=1b5-03}, we can confirm that signal of crossover about $\sigma_4$ 
is exactly same as that of an ordinary four-dimensional lattice gauge theory ($\beta_5=0$).

\begin{figure}[p]
\begin{center}
\includegraphics[scale=0.4]{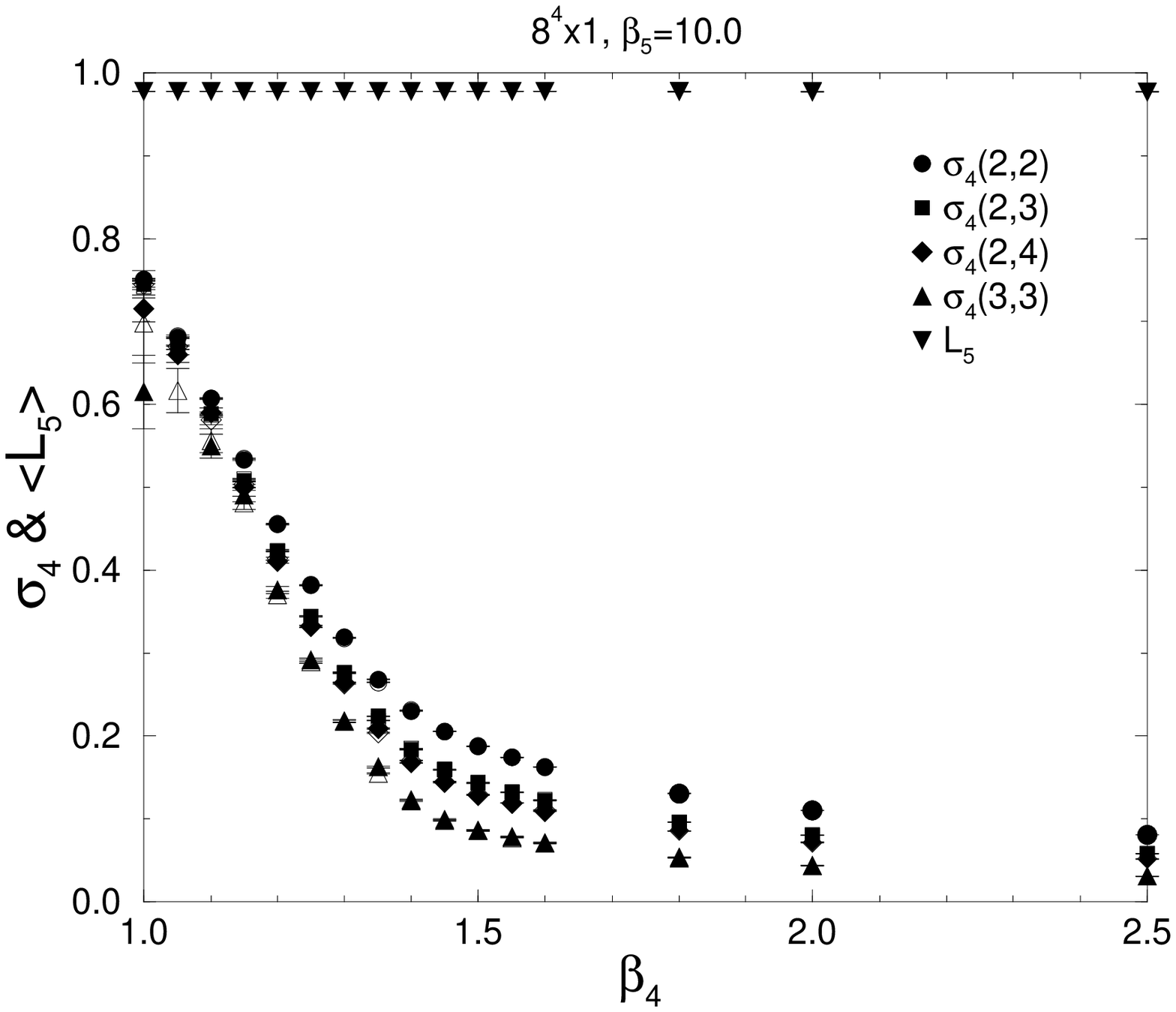}
\vspace{5cm}
\caption{$\sigma_4$ and $\kbra{L_5}$ at $\beta_5=10.0$. }
\label{fign5=1b5-10}
\end{center}
\end{figure}

\vspace{1mm}
(iv) ${\it \beta_5 =10.0}$:
We can see that $\sigma_4$ is decreasing as $\beta_4$ is growing in Fig.\ref{fign5=1b5-10},
although it is difficult to recognize a rapid change of  $\sigma_4$ at $\beta_4 \sim 1.2$. 

The naive continuum limit of an $N_5=1$ system corresponds to the four-dimensional
 gauged non-linear $\sigma$ model:
\beq
S_{\rm lat}
\rightarrow 
\int d^4 x \left\{
\frac{1}{4} \Tr F_{\mu\nu}^2 + \frac{1}{2} f_{\pi}^2 \Bigl(D_{\mu}V(x) \Bigr)^{\dagger} \Bigl( D_{\mu}V(x) \Bigr)
\right\}
\label{gnlsmact}
\eeq
where $V(x)$ is a scalar field with a constraint condition 
\beq
&& V(x)
= \theta_0(x)+i\vec{\sigma}\cdot\vec{\theta}(x) \ ,\nn\\
&& \quad \theta_0^2+\vec{\theta}^2=1\ .
\eeq
This condition reads to $(\del_{\mu}\theta_0)^2 \ll (\del_{\mu}\theta^a)^2 $
at $\theta^a \ll 1$.
Then the system corresponds to a Georgi-Glashow model  with a Higgs triplet
\cite{B-G-L86,L-S86,B-R87} which is equivalent to a four-dimensional $O(3)$ 
Heisenberg model. 


\subsection{Case of $\bm{N_5>1}$} 

\begin{figure}[p]
\begin{center}
\includegraphics[scale=0.5]{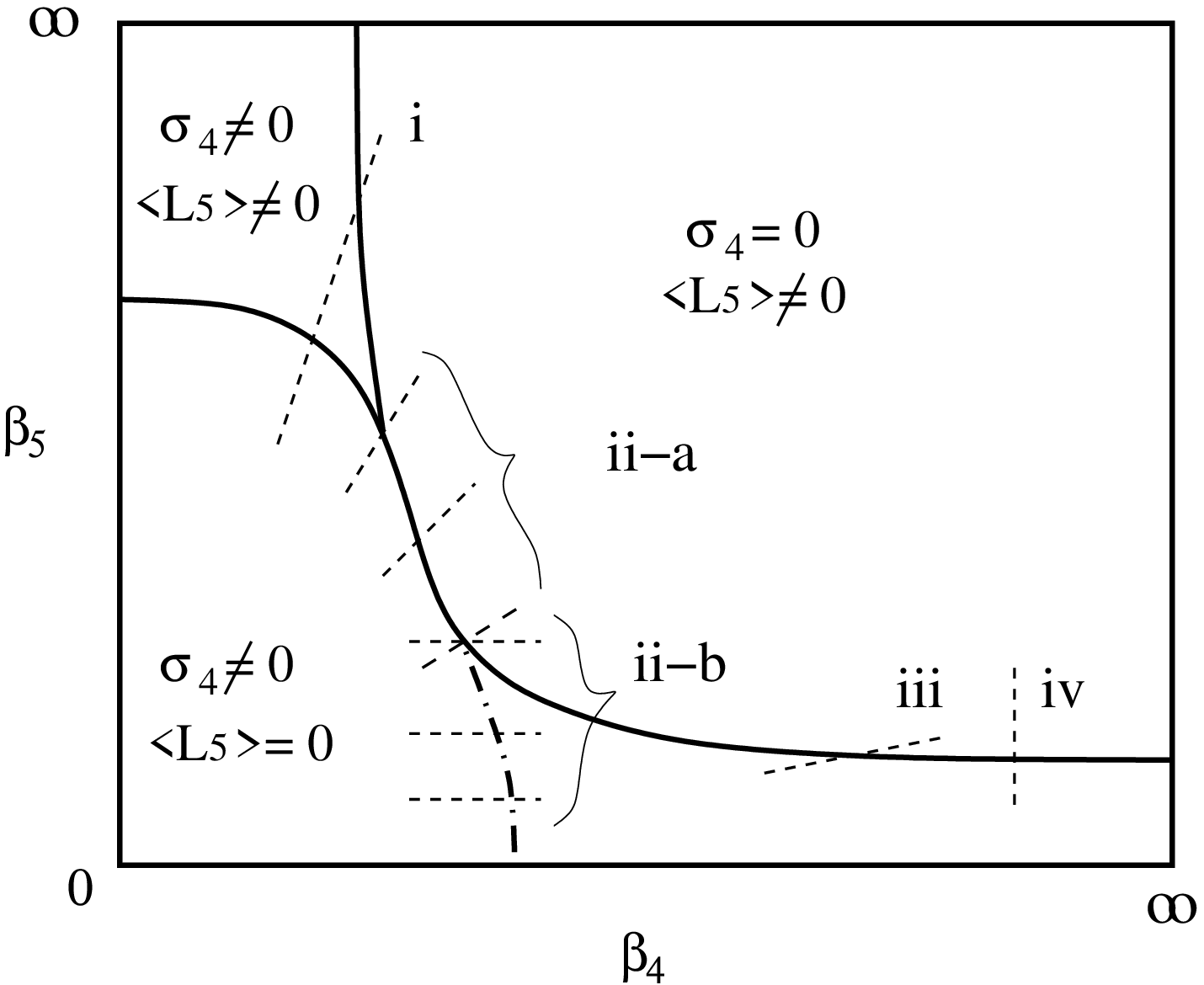}
\vspace{5cm}
\caption{
The phase diagram for $N_5 > 1$. The meaning of lines is same as Fig.3.}
\label{phdiagN5=2up}
\end{center}
\end{figure}

The phase diagram for $N_5>1$ and measuring directions
 are shown in Fig.\ref{phdiagN5=2up}.
Dashed lines are  
 (i) $\gamma=\sqrt{2}$, (ii) $\gamma = 0.5,~1/\sqrt{2}~(\sim 0.71),~1/1.2~(\sim 0.83)$,
 (iii) $\gamma=0.29$ and (iv) $\beta_4=8.0,~10.0$ and $20.0$ respectively.   \\

\begin{figure}[p]
\includegraphics[scale=0.4]{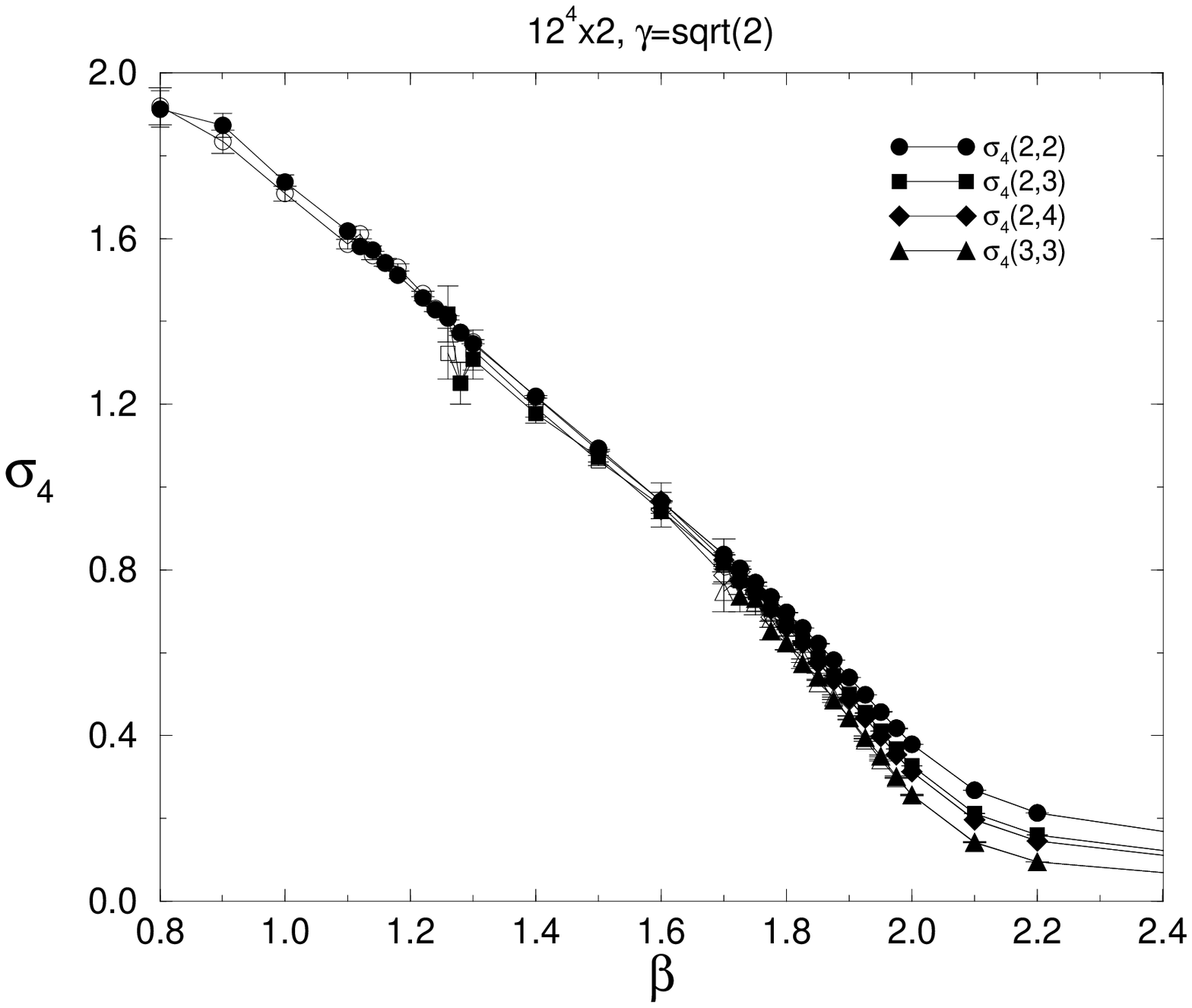}
\includegraphics[scale=0.4]{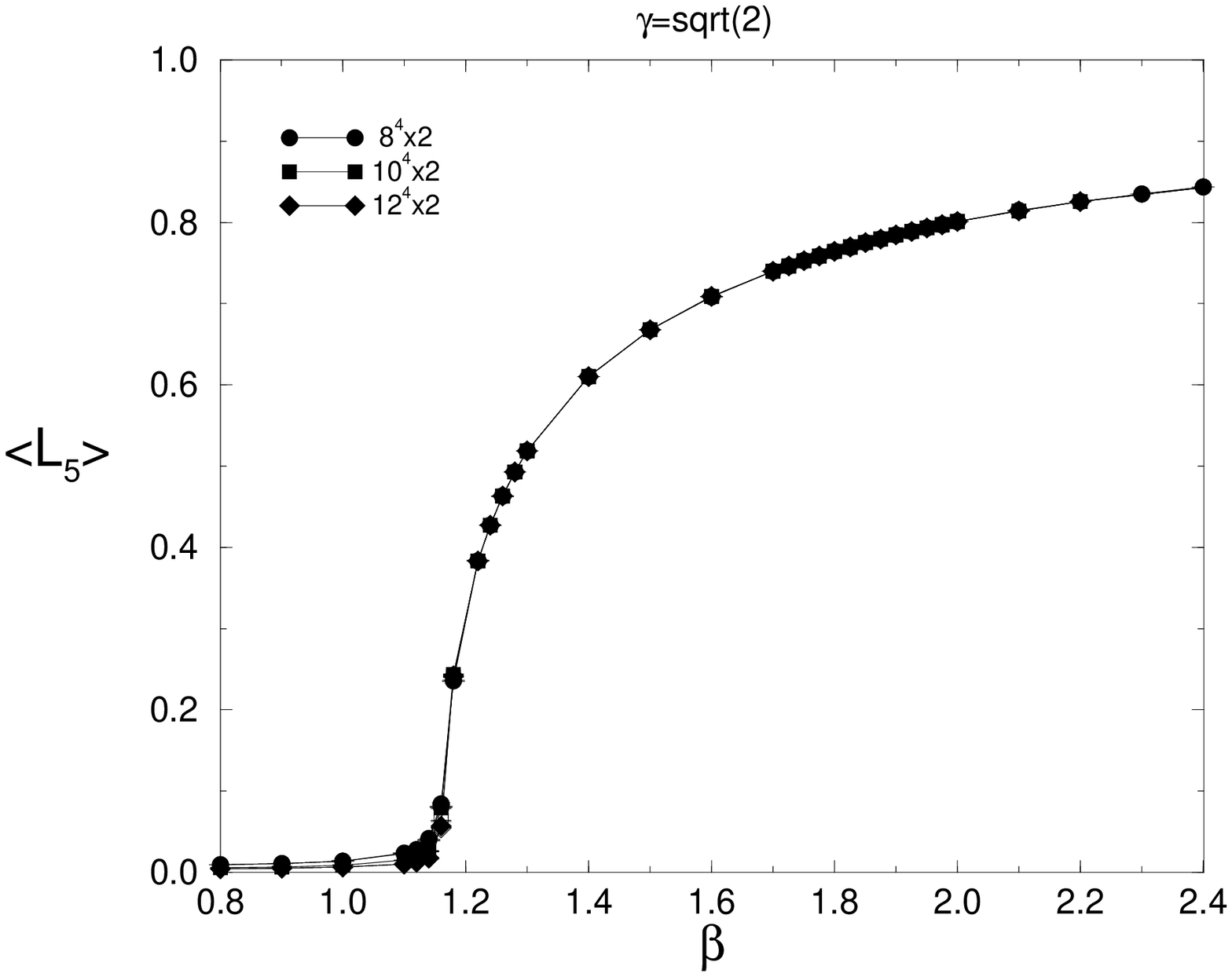}
\vspace{5cm}
\caption{$\kbra{L_5}$ and $\sigma_4$ at $\gamma=\sqrt{2}$. }
\label{fign5=2gr2}
\end{figure}

(i) ${\it \gamma=\sqrt{2}}$: 
In Fig.\ref{fign5=2gr2}, as $\beta$ becomes to large,
 the new phase appears at $\beta \sim 1.2$ where the phase varies
 from $\sigma_4 \neq 0$ and $\kbra{L_5}=0$
 to $\sigma_4 \neq 0$ and $\kbra{L_5} \neq 0$.
Afterward the new phase transition occurs at $\beta \sim 2.0$
 where the phase varies from $\sigma_4 \neq 0$ and $\kbra{L_5} \neq 0$
 to $\sigma_4 = 0$ and $\kbra{L_5} \neq 0$. 

\begin{figure}[p]
\includegraphics[scale=0.4]{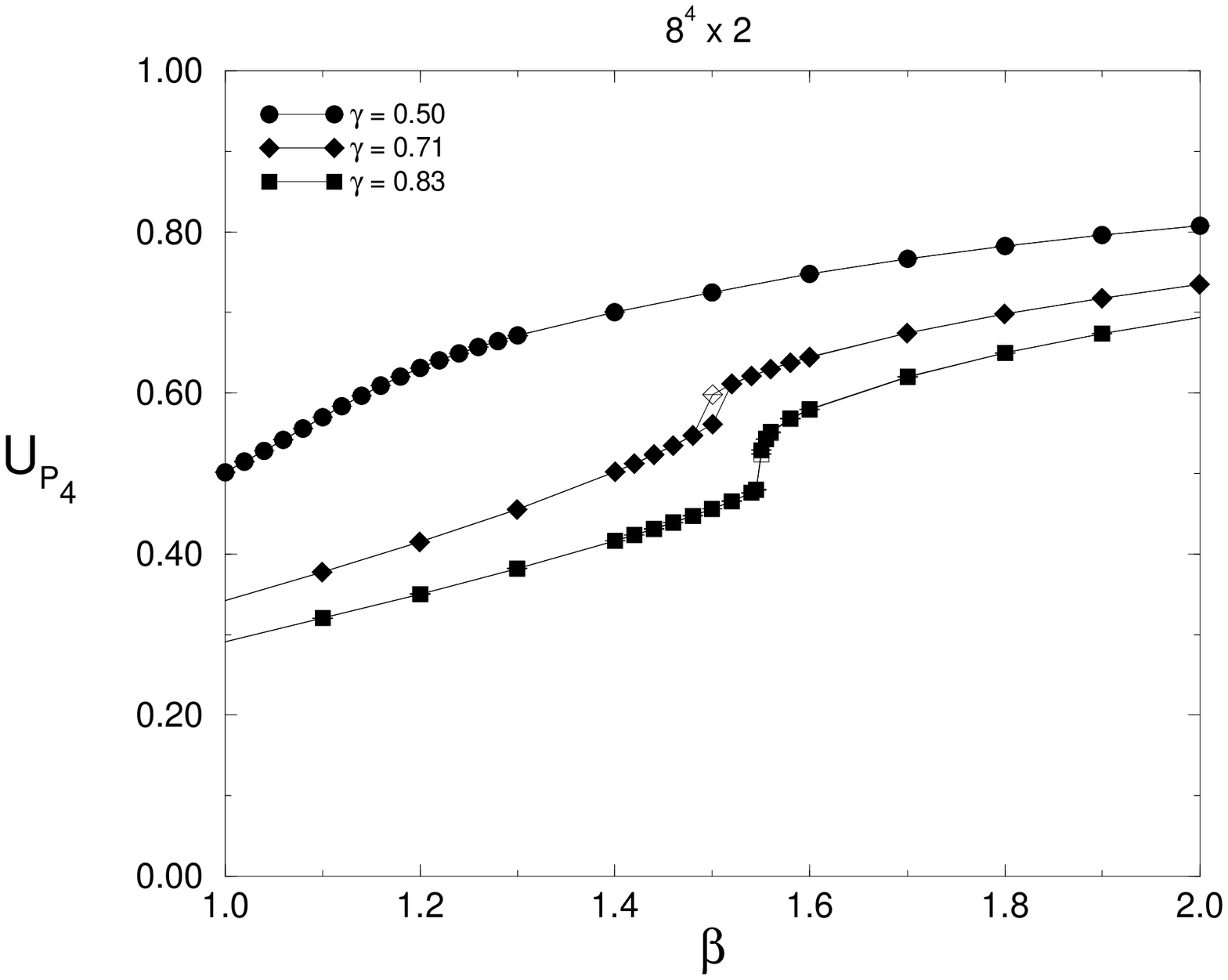}
\includegraphics[scale=0.4]{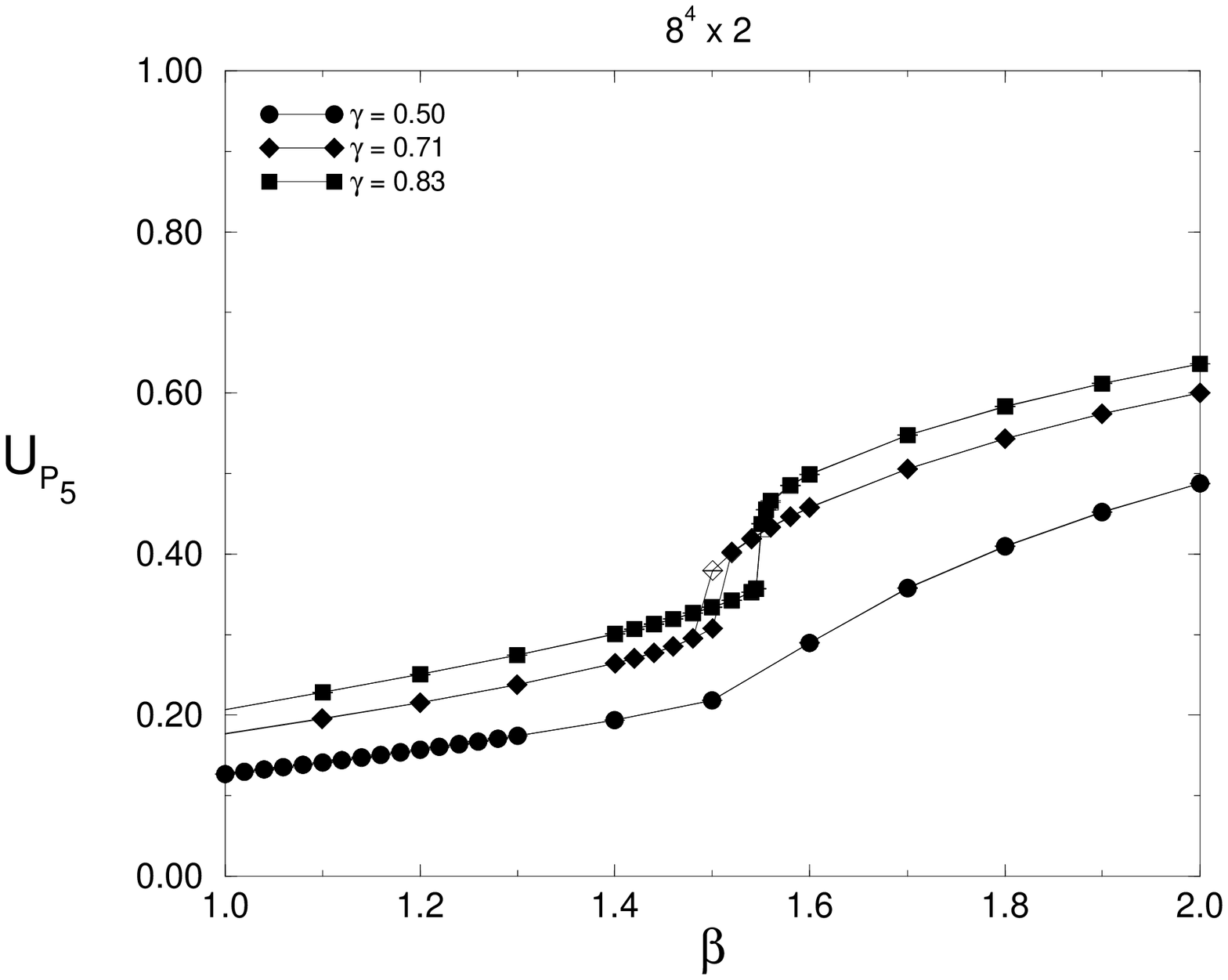}
\vspace{5cm}
\caption{$U_{P_4}$ and $U_{P_5}$ at $0.5 \leq \gamma \leq 0.83$.}
\label{fign5=2g050-083}
\end{figure}

\vspace{1mm}
(ii-a) ${\it 0.5 \leq \gamma \leq 0.83 }$: 
In the middle range of $\gamma$,
 we found a first order phase transition for 
both order parameters $U_{P_4}$ and  $U_{P_5}$ on a $8^4 \times 2$ lattice 
in Fig.\ref{fign5=2g050-083}.
There the hysteresis loop is clearly seen at $\gamma = 0.71$.
As $N_5$ approaches to $N_4$, the range is wider and 
includes  unity where the theory is recovered as usual five-dimensional one.
Then this signal of the first order transition corresponds to that of
 the isotropic five-dimensional $SU(2)$ gauge theory \cite{K-N-O92}. 

\begin{figure}[p]
\begin{center}
\includegraphics[scale=0.4]{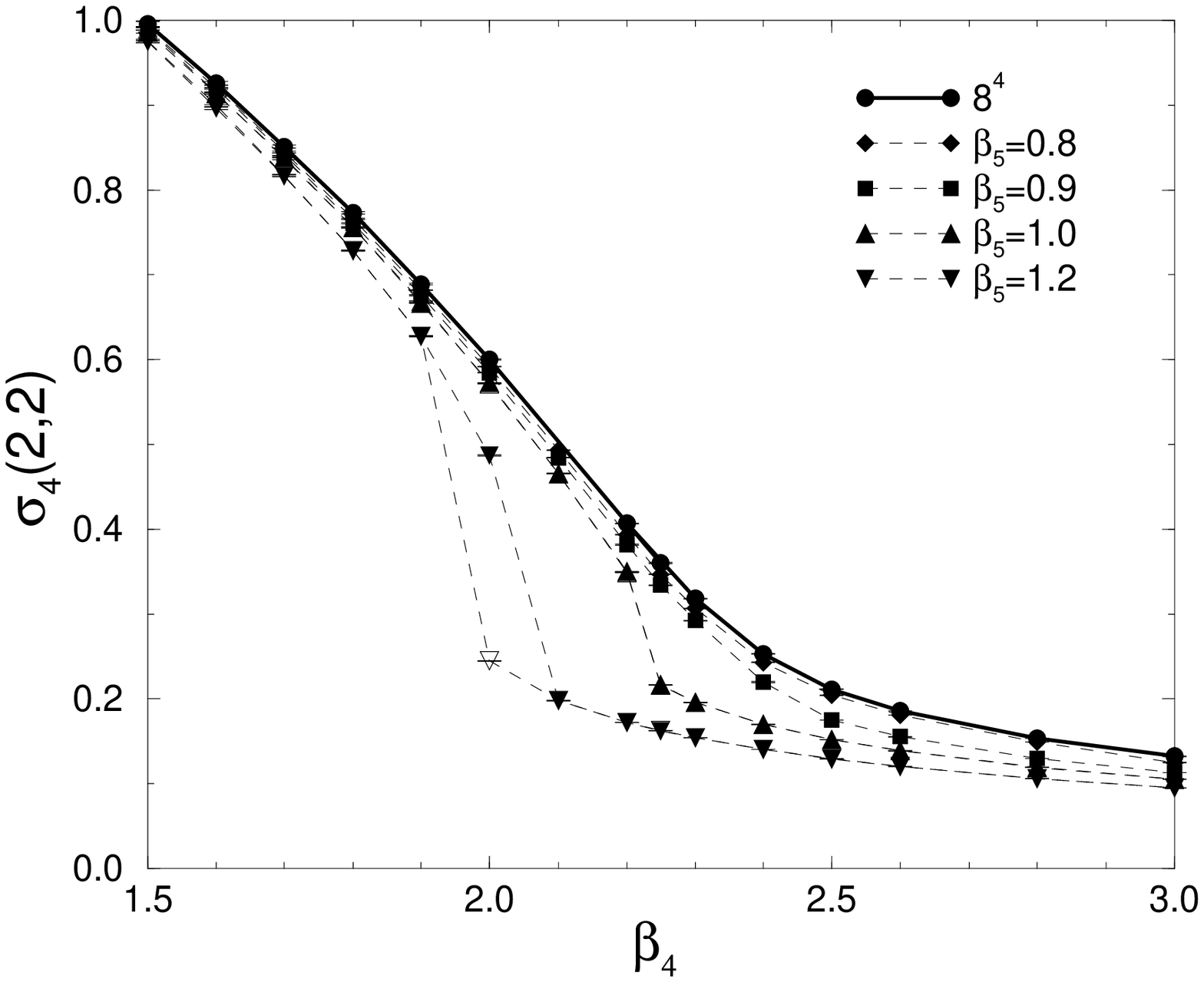}
\vspace{5cm}
\caption{The circle symbol shows $\sigma_4(2,2)$ on an $8^4$ lattice.
The others are results on an $8^4 \times 3$ lattice.}
\label{fign5=3b5-00-10}
\end{center}
\end{figure}

\begin{figure}[p]
\begin{center}
\includegraphics[scale=0.4]{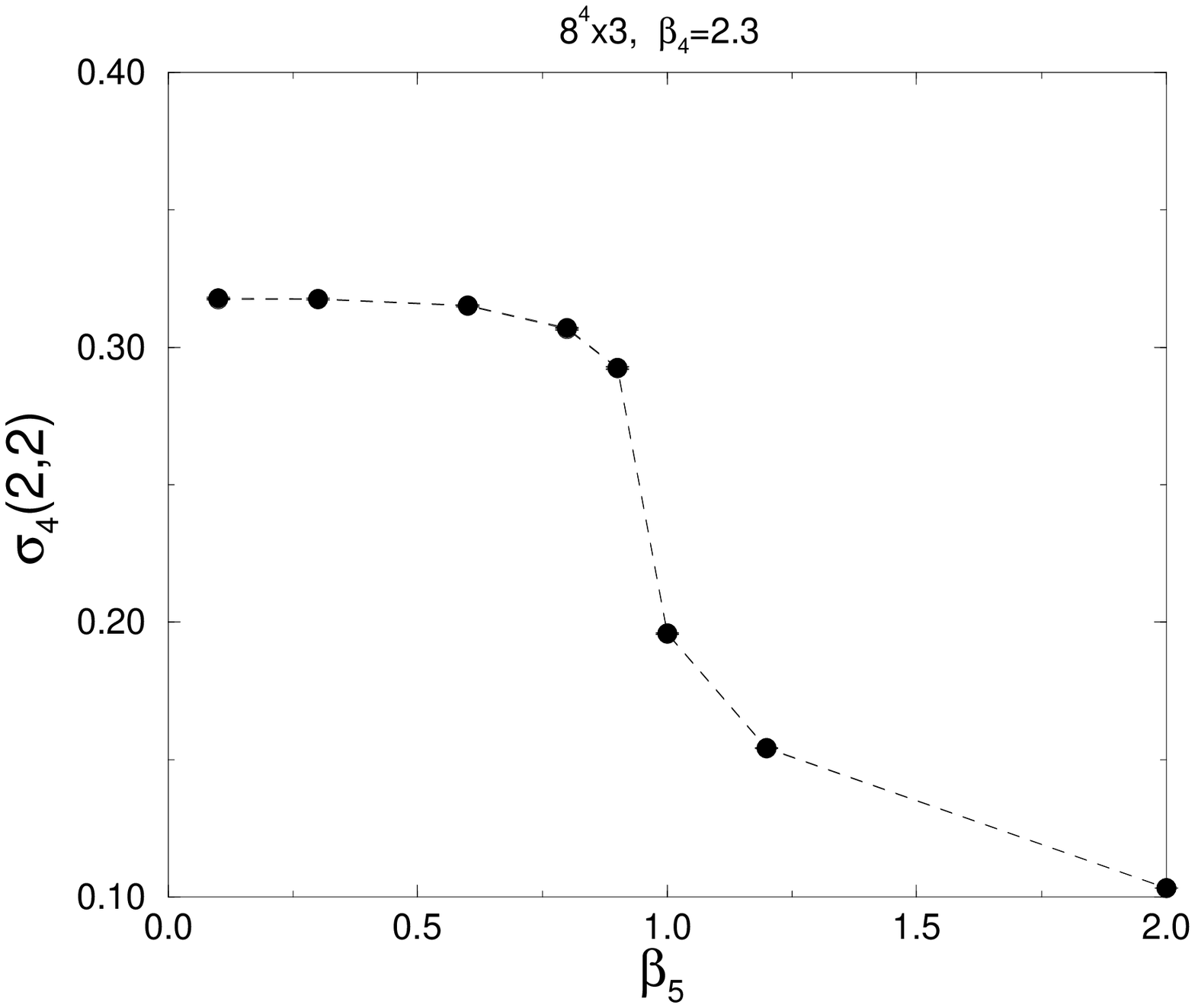}
\vspace{5cm}
\caption{$\sigma_4(2,2)$ at $\beta_4 = 2.3$.}
\label{fign5=3b4-230}
\end{center}
\end{figure}

\vspace{1mm}
(ii-b) ${\it 0.0 \leq \beta_5 \leq 1.2}$: 
Fig.\ref{fign5=3b5-00-10} shows a change of $\sigma_4(2,2)$ for $\beta_4$. 
To compare with it, the result of an ordinary four-dimensional lattice is also shown
 at the circle symbol.
As $\beta_5$ is larger, a slope of $\sigma_4(2,2)$ is deeper.
The signal of crossover is changed to the second order 
phase transition at $\beta_5 \sim 1.0$ and further the first order phase transition 
at $\beta_5 \sim 1.2$.
The change of $\sigma_4(2,2)$ for $\beta_5$ at $\beta_4=2.3$, 
is shown in Fig.\ref{fign5=3b4-230}.
We can see that  $\sigma_4(2,2)$ or the tension is slightly varied in small $\beta_5$, 
although it becomes clear in $\beta_5 > 0.8$ near the phase transition. 

\begin{figure}[p]
\includegraphics[scale=0.4]{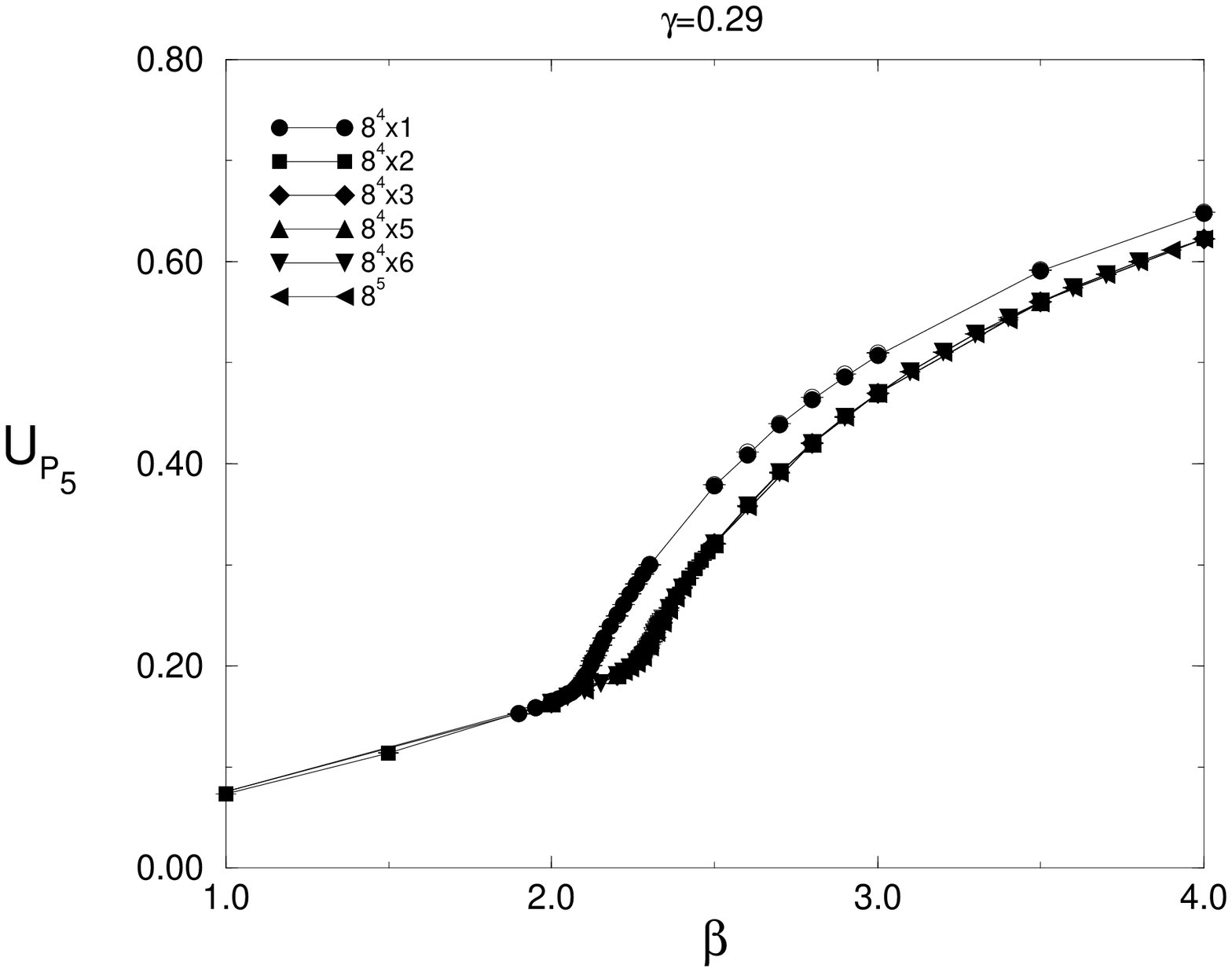}
\includegraphics[scale=0.4]{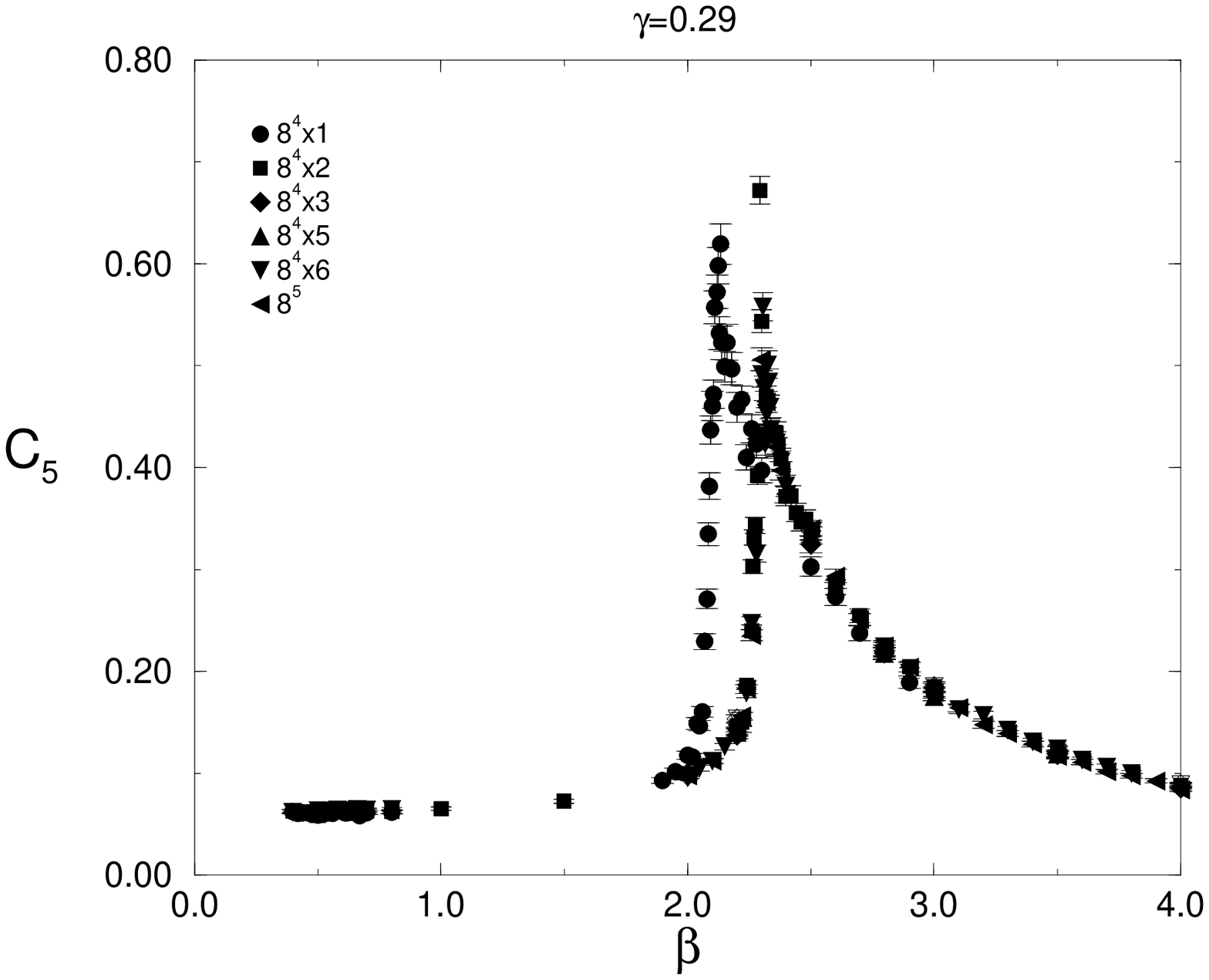}
\vspace{5cm}
\caption{$U_{P_5}$ and $C_5$ at $\gamma=0.29$ on an $8^4\times N_5$ lattice; 
$N_5=1,2,3,5,6$ and $8$.}
\label{fign5=1-8g028}
\end{figure}

\vspace{1mm}
(iii) ${\it \gamma=0.29}$: 
We can find the phase transition at $\beta \sim 2.3$ 
($\beta_4 \sim 0.79, \beta_5 \sim 0.67$).
It is a surprising fact in a multi-layer world 
 that the transition point is independent of $N_5 >1$.
Fig.\ref{fign5=1-8g028} shows results of $U_{P_5}$ and $C_5$ on
 a $8^4\times N_5$ lattice, where $N_5=1,2,3,5,6$ and $8$.
Both quantities for $N_5 >1$ are unchanged. 

\begin{figure}[p]
\includegraphics[scale=0.4]{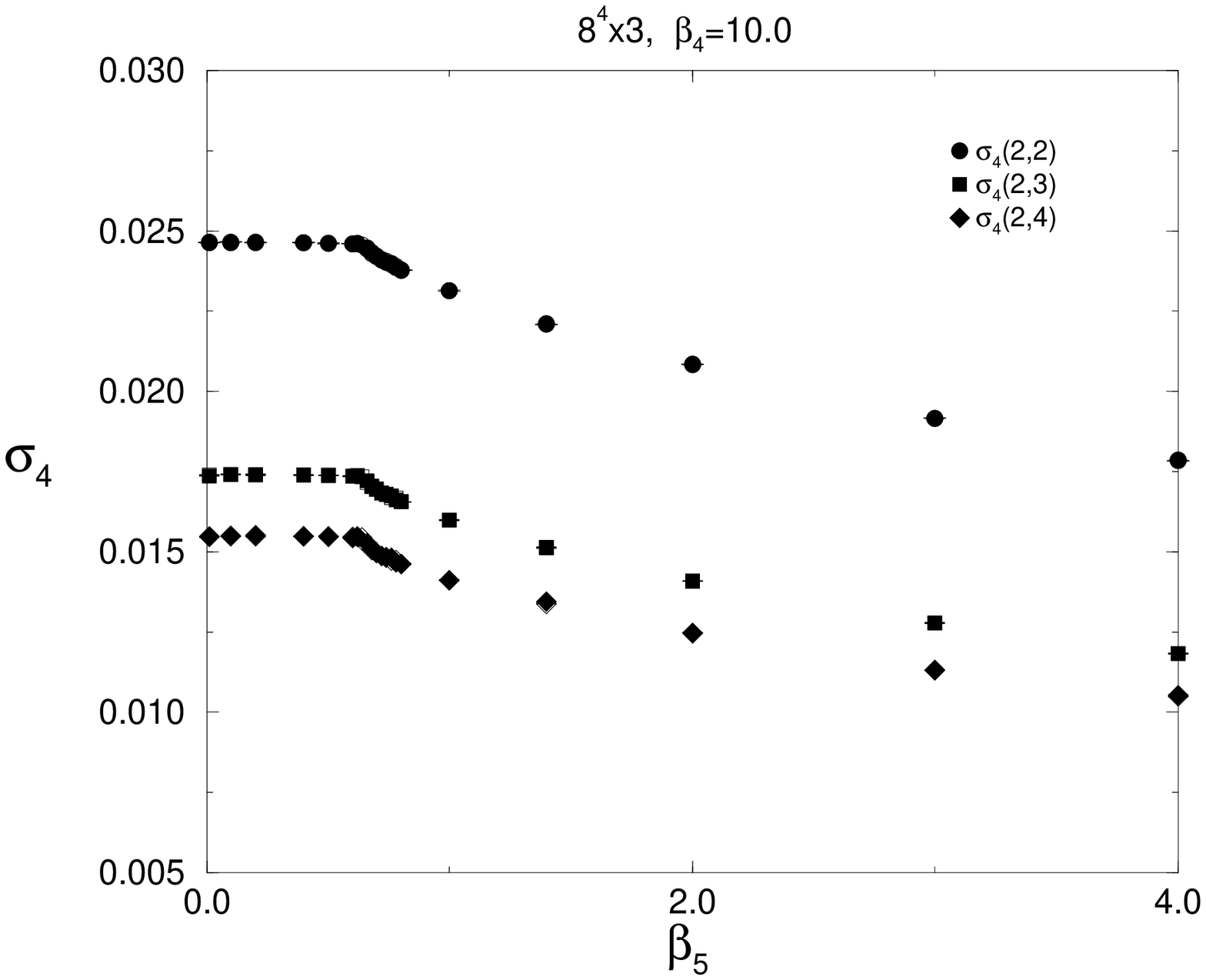}
\includegraphics[scale=0.4]{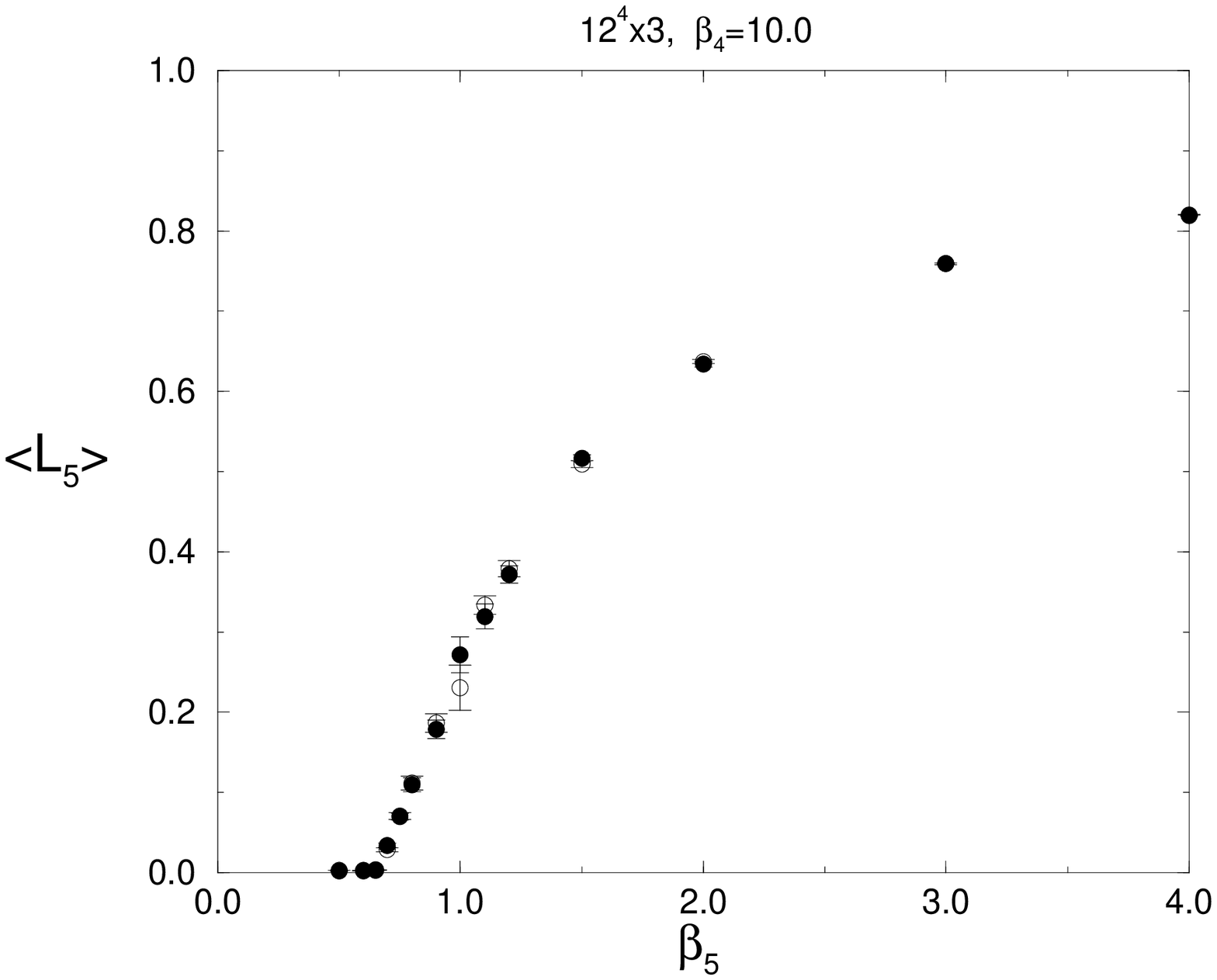}
\vspace{5cm}
\caption{$\kbra{L_5}$ and $\sigma_4$ at $\beta_4=10.0$.}
\label{fign5=3b4-10}
\end{figure}

\begin{figure}[p]
\begin{center}
\includegraphics[scale=0.45]{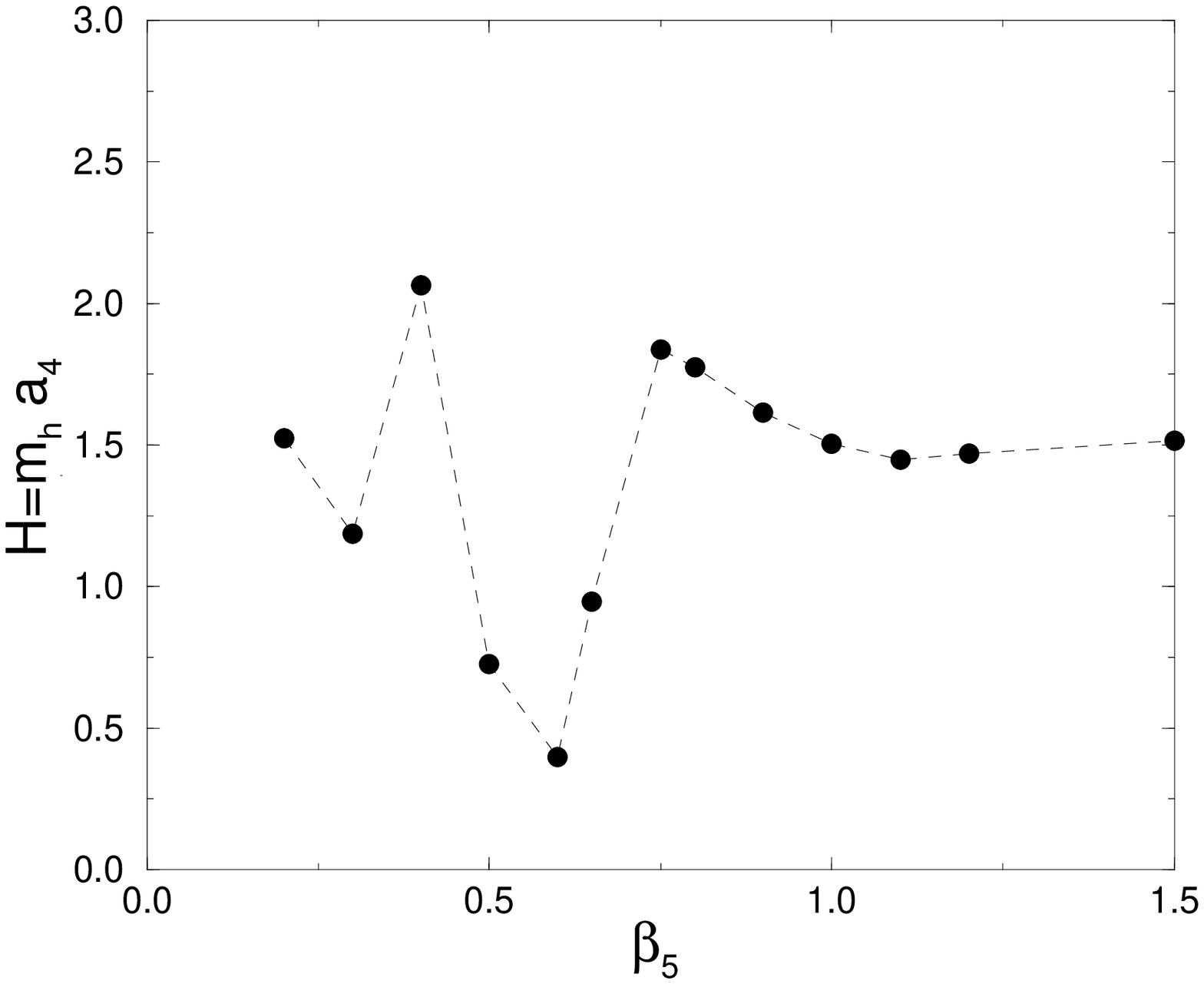}
\vspace{5cm}
\caption{The lattice counterpart $H$ for a Higgs mass at $\beta_4=10.0$
 on a $12^4\times 3$ lattice.}
\label{figK}
\end{center}
\end{figure}

\begin{figure}[p]
\includegraphics[scale=0.4]{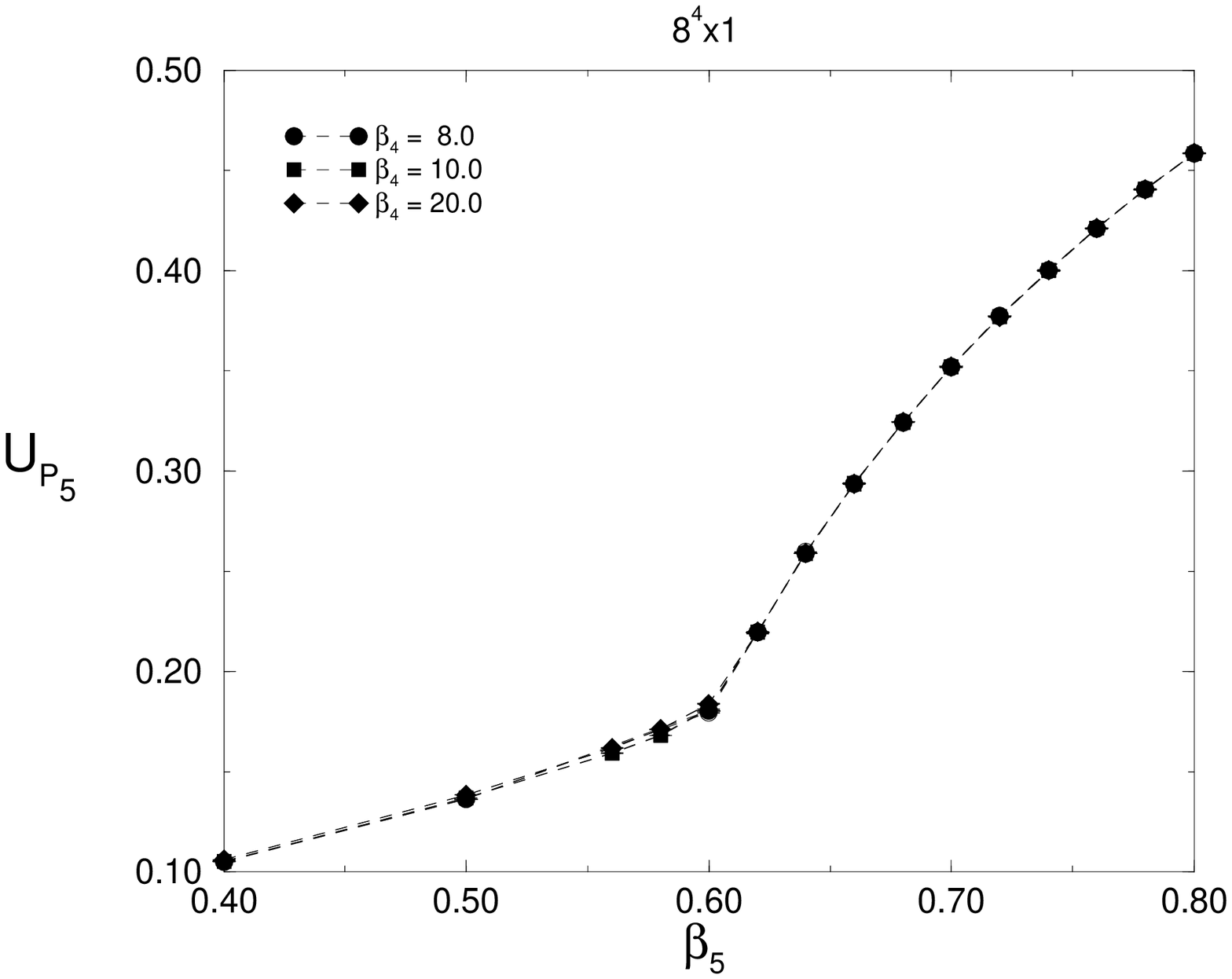}
\includegraphics[scale=0.4]{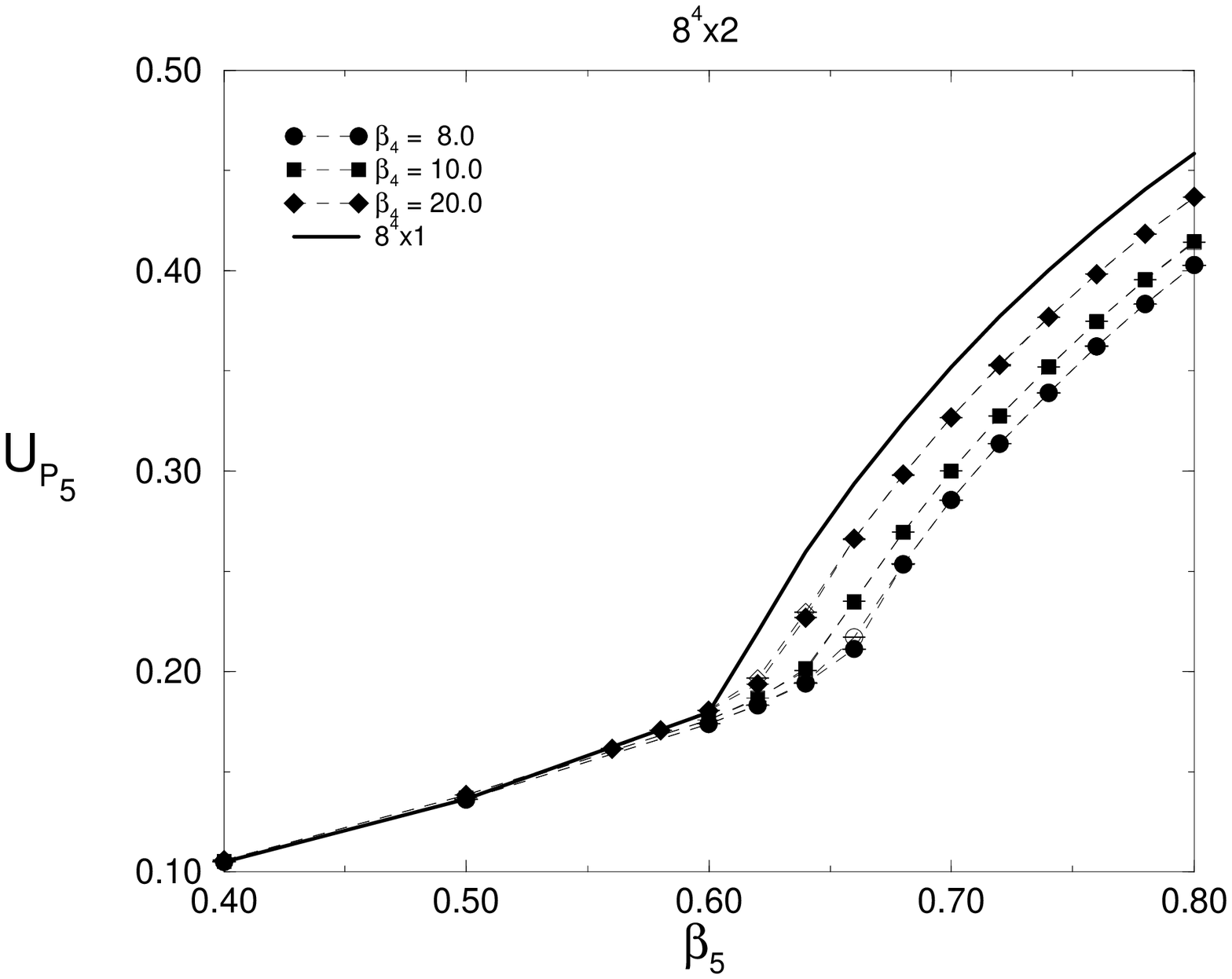}
\vspace{5cm}
\caption{$U_{P_5}$ at $\beta_4=8.0,10.0$ and $20.0$. }
\label{fign5=1-3b4-8-20}
\end{figure}

\vspace{1mm}
(iv) ${\it \beta_4=8.0,10.0}$ {\it and} ${\it 20.0}$: 
At $\beta_4=10.0$ we can see that the phase transition occurs at
 $\beta_5 \sim 0.63$  from $\sigma_4 \neq 0$ and $\kbra{L_5}=0$ to $\sigma_4=0$ and
 $\kbra{L_5}\neq 0$ $-\!\!\!-$ see Fig.\ref{fign5=3b4-10}.
The hysteresis loop can not be seen in their signals.
Therefore we expect the phase transition  is second order.
The inverse correlation length for Polyakov loops shows at $\beta_4=10.0$ on 
a $12^4\times 3$ lattice in Fig.\ref{figK} and also supports the property of the phase transition.
In addition, we found that the transition point for $N_5>1$ is closer to that 
for $N_5=1$ as $\beta_4$ is larger $-\!\!\!-$ see Fig.\ref{fign5=1-3b4-8-20}.
 

Those interesting phenomenon in the multi-layer world shall support our discussions in
 the next section. 


\section{Multi-Layer World} 

\subsection{Multi-layer world and gauge fields}

In this section, we investigate a multi-layer world 
 ($\gamma$ is sufficiently  small to unity and $N_5 >1$). 
This world is composed of four-dimensional gauge theories bound by `scalar' fields:
$V_n^i \equiv U_{n,5}^i$.
The fields are transformed as bi-fundamental representations
of ${\cal L}_i$ and ${\cal L}_{i+1}$.
The gauge transformations are defined as 
\beq
U_{n,\mu}^{i} &\rightarrow& Y^{i}(n)\ U_{n,\mu}^{i}\ {Y^{i}}^{\dagger}(n+\hat{\mu})
\hspace{0.4cm} \mbox{for a gauge field inside a ${\cal L}_i$, } \\
V_{n}^{i} &\rightarrow& Y^{i}(n)\ V_{n}^{i}\ {Y^{i+1}}^{\dagger}(n)
\hspace{1cm} \mbox{for a `scalar' field between ${\cal L}_i$ and ${\cal L}_{i+1}$, }
\label{gaugetr}
\eeq
where $Y^i(n)$ is an arbitrary element of $SU(2)$.
From (\ref{gaugetr}), $V_n^i$ can be fixed to unity except for $V_n^1$,
\beq
\left\{
\begin{array}{lclll}
V_{n}^{1'}
& = & Y^{1}(n)\ V_{n}^{1}\ {Y^{2}}^{\dagger}(n)& = 1\ ,&\\
V_{n}^{2'}
& = & Y^{2}(n)\ V_{n}^{2}\ {Y^{3}}^{\dagger}(n)& = 1\ ,& \\
&\vdots& \\
V_{n}^{N_5'}
& = & Y^{N_5}(n)\ V_{n}^{N_5}\ {Y^{1}}^{\dagger}(n)& =Y^{1}(n)\left\{\prod_{i=1}^{N_5} V_{n}^{i}\right\}{Y^{1}}^{\dagger}(n) &\equiv Y^1(n)X_n {Y^{1}}^{\dagger}(n)\ .
\end{array}\right. 
\label{gaugefix}
\eeq
Note that  $\Tr V_{n}^{N_5'}$ is exactly a Polyakov loop $L_{5,n}$.
Since a non-trivial $V_n^i$ can be moved to every inter-layer, this implies 
 that the system has a translational invariance to the fifth direction.
In this point, the multi-layer world is significantly different from
 a just four-dimensional theory.

Based on this setting of the  gauge fixing, we consider mass spectrum of 
vector fields, $A_{n,\mu}^{a\ i}$.  
The masses include the effect of fluctuations by layers.  
We get the following mass term,
\beq
\mbox{($A_{n,\mu}^2$ term)}
&\equiv&
a_4^4\sum_{n,\mu}\sum_{i,j=1}^{N_5} 
\frac{1}{2} A_{n,\mu}^{a\ i}\ M_{ij}^2\ A_{n,\mu}^{a\ j}\ ,
\eeq
\noindent
where the mass squared matrix is defined as 
\beq
M_{ij}^2
&\equiv& \frac{\beta_5 g_4^2 }{4\ a_4^2} \left\{
   \left[
    \begin{array}{ccccc}
    1 & -1 &        & & \\
    -1&  2 & \ddots & &  \\
      & \ddots & \ddots & \ddots & \\
      &    & \ddots & 2 & -1 \\
      &    &        & -1&  1 \\
    \end{array}
    \right]
   + L_{5,n}^2
   \left[
    \begin{array}{ccccc}
    1 &  &     &   &  -1\\
      & 0&     &  0& \\
      &  & \ddots &   &  \\
      & 0&        & 0 &  \\
    -1&  &        &   &  1 \\
    \end{array}
    \right]
   \right\} \nn \\
&\equiv& \frac{\beta_5 g_4^2 }{4\ a_4^2} {\cal M}_{ij} 
\label{kkmass}
\eeq
using $U_{n,\mu}^{i} = \exp\left\{i\ a_4\ g_4\ \frac{\sigma^a}{2}\ 
 A_{n,\mu}^{a\ i}\right\}$ .
For the purpose that we go ahead to non-perturbative analysis, 
the mass may be also non-perturbatively defined by the correlation length 
between different layers. But the analysis is necessary for  
a large $N_5$ lattice and it is beyond our present computer capacity. 
Therefore, we take the semi-classical approach such  that the mass formula 
is derived by expanding a field $A_{n,\mu}$  and the mass squared matrix is calculated 
by Monte Carlo simulation.
Under this approximation,  $L_{5,n}$  can be replaced with  
$\kbra{L_{5}} \equiv 1/(2V_4) \cdot  \sum_n \Tr X_n$, and $V_4$ is the 
four-dimensional volume of a layer. In the case of $\kbra{L_5}^2=1$, 
we can easily analyze this  matrix ${\cal M}_{ij}$  owing to
 the periodic boundary condition for the fifth direction \cite{A-C-G01,C-H-P-W01}. 
Eigenvalues of this matrix are
\beq
\lambda_k
&=& 4 \sin^2\frac{\pi k}{N_5 }
\quad,\quad k= 0,1,\cdots, N_5-1 \ ,
\eeq
and at $k \ll N_5$, they are simply written as
\beq
\lambda_k = \left(\frac{2\pi k}{N_5}\right)^2\ .
\label{lambdak}
\eeq

\begin{figure}[p]
\hspace{-0.05\textwidth}
\begin{minipage}{0.3\textwidth}
\includegraphics[scale=0.3]{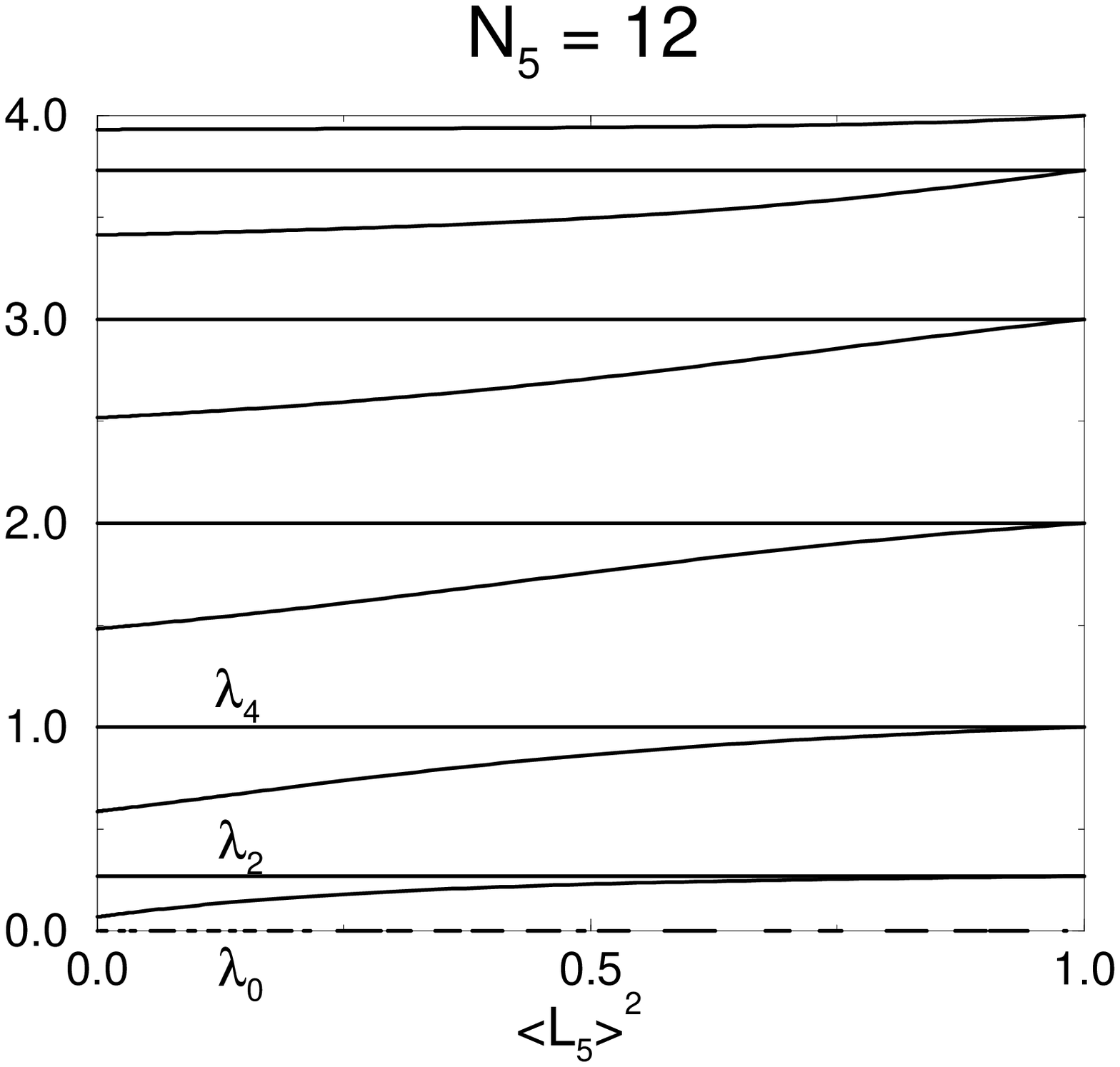}
\end{minipage}
\hspace{0.02\textwidth}
\begin{minipage}{0.3\textwidth}
\includegraphics[scale=0.3]{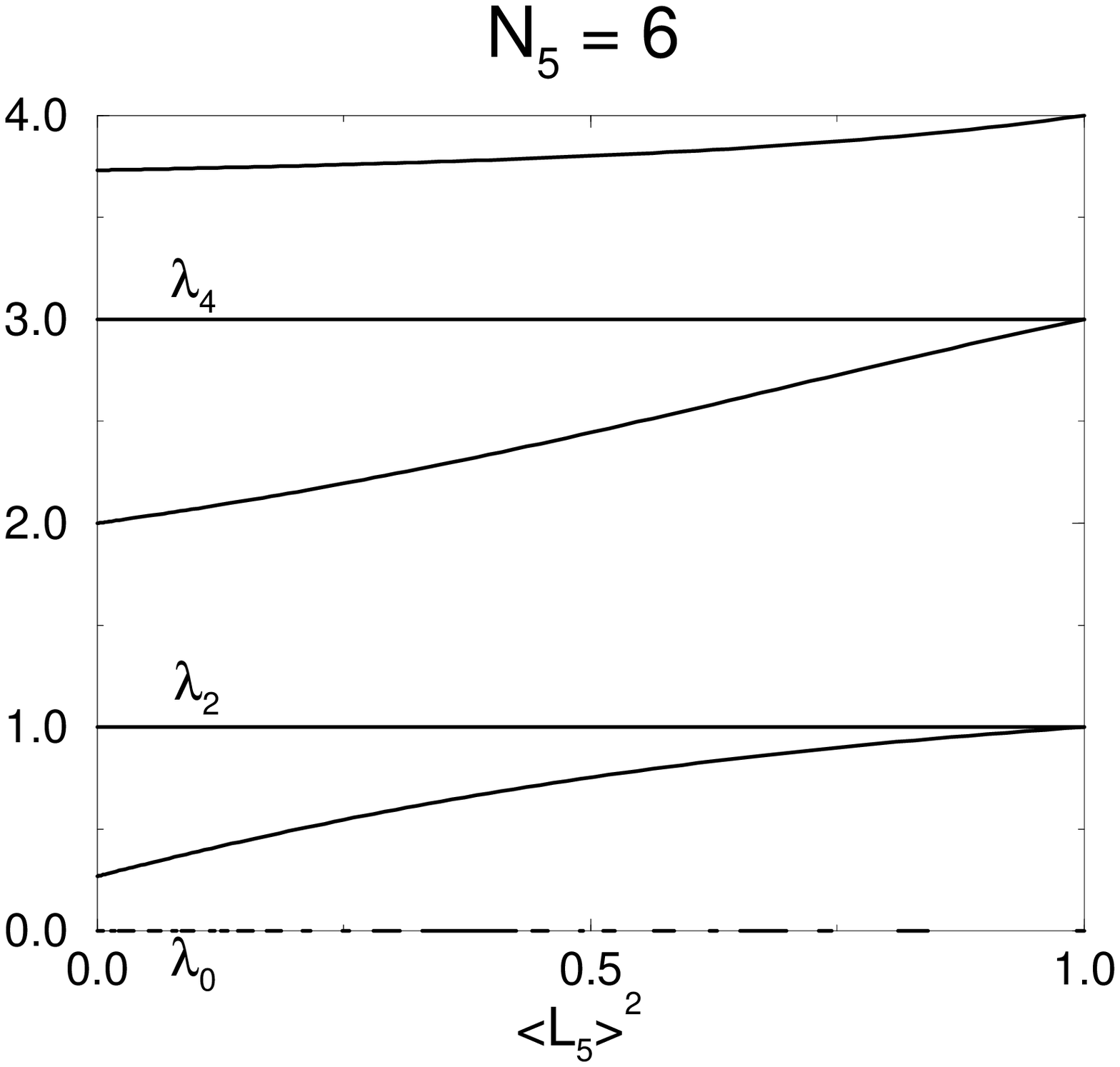}
\end{minipage}
\hspace{0.02\textwidth}
\begin{minipage}{0.3\textwidth}
\includegraphics[scale=0.3]{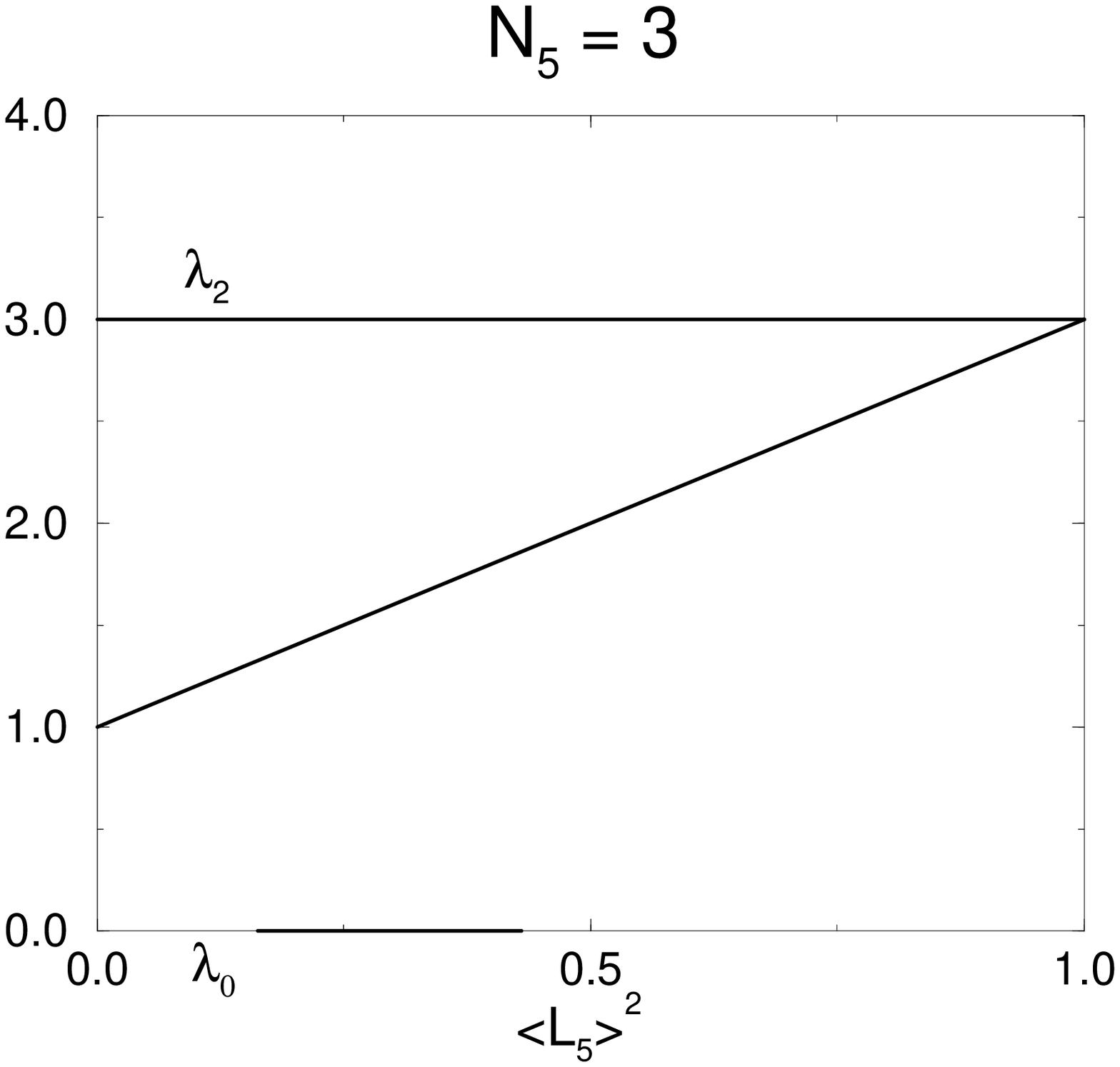}
\end{minipage}
\vspace{5cm}
\caption{Eigenvalues $\lambda_k$ of a matrix ${\cal M}_{ij}$ for 
$N_5=12,6$ and $3$.}\label{lambda-L5}
\end{figure}

It is noted that the values for even $k$ are independent of $\kbra{L_5}$. 
Since it is not so easy to get the values for odd $k$, we numerically 
calculate them with arbitrary $\kbra{L_5}$ which is determined by Monte Carlo 
simulation. 
Fig.\ref{lambda-L5} shows plots of eigenvalues $\lambda_k$ versus $\kbra{L_5}^2$ 
 at $N_5=3,6$ and $12$.


We see that eigenvalues are doubly degenerate at $\kbra{L_5}^2=1$.
In addition, a noteworthy thing is that the eigenvalues at $\kbra{L_5}^2=1$ and  
$N_5=n$ are equal to the eigenvalues at $\kbra{L_5}=0$ and $N_5=n/2$ 
except for its degeneracy. 
Our system at $\kbra{L_5}^2=1$ and  $N_5=n$  has  really the  corresponding 
relation to  a circular model at $N_5=n$,  and  a system with $\kbra{L_5}=0$ and 
$N_5=n/2$  has a relation to a model with $Z_2$ orbifolding \cite{A-C-G01,C-H-P-W01}. 
This circumstance means that $S^1$ and $S^1/Z_2$ with respectively different
 topologies are connected by one parameter $\kbra{L_5}$
 which corresponds to the energy releasing quarks from a layer. 
The reason is the existence of 'singular' 
gauge transformations  such as  (\ref{gaugefix})
on lattice theories in contrast to continuum theories. 
The eigenvalues for even $k$ are topology-independent and 
mainly used in our following analysis 
because we do not have a clue to change to the topology 
 although topology changing phenomena is an interesting  topics.

\subsection{Continuum limit as a four-dimensional gauge theory} 

In this subsection, we shall consider the continuum limit of four-dimensional 
gauge theories   on  layers.  Since our fifth direction is identified with 
just a internal space, we  take only the continuum limit $a_4 \rightarrow 0$ 
on layers.  From  $\beta_4 = 4/ g_4^2 $, we can express the mass squared 
matrix as

\beq
M_k^2 = \frac{\beta_5 }{\beta_4 a_4^2}\lambda_k \left(\kbra{L_5(\beta_4,\beta_5)},N_5 \right)
= \frac{\gamma^2}{a_4^2}\lambda_k(\kbra{L_5(\beta_4,\beta_5)},N_5)\ .
\label{fourmass}
\eeq
\noindent
At $k \ll N_5$, the simple mass formula
\beq
M_k^2 = \gamma^2 \left(\frac{2\pi k}{N_5 a_4} \right)^2 =  
\frac{1}{N_5} \left(\frac{2\pi \gamma k}{a_4} \right)^2\ ,
\label{kkm}
\eeq
is obtained.
In order to take the continuum limit $a_4\rightarrow 0$ with 
a fixed mass (\ref{fourmass}), a possible method exists;
\noindent
$$
N_5 {\rm ~fixed,~} \gamma \rightarrow 0 {~\rm and ~} \frac{\gamma}{a_4}\ 
{~\rm finite ~}.
$$ 
\noindent
The approach is realized on $\beta_4 \rightarrow \infty$  
with finite $\beta_5$. 
This theory is defined as four-dimensional asymptotically free one 
using $\sigma_4$ in the confinement phase like a la deconstruction. 
Actually, the scaling for $\sigma_4$ is non-trivial because we have to 
take care of bi-fundamental fields $V_n^{n_5}$. 
Our numerical result shows that the scaling is subtly different 
from an ordinary four-dimensional theory with $\beta_5=0$ $-\!\!\!-$ see 
Fig.\ref{fign5=3b5-00-10}.

Although  we can not use  $\sigma_4$ in the case of the deconfinement phase, 
from (\ref{Kdef}) Higgs mass, 
\beq
m_h \equiv \frac{H(\beta_4,\beta_5,N_5)}{a_4}\ ,
\label{Hdef}
\eeq
can be considered instead. 
This quantity is not yet seen as a good data in statistics in 
Fig.\ref{figK}. 
But near $\beta_{5c}$ in both phases, it can be seen as 
the important scaling parameter
\beq
H \propto |\beta_5 - \beta_{5c}|^{\nu}
\label{crit-nu}
\eeq
\noindent
where $\nu$ is a critical exponent for the correlation length 
and seems less than unity. 
It is consistent with second order phase transition 
at $\beta_5 = \beta_{5c} \sim 0.6$. 
The observation is supported by no hysteresis loop 
and the peak of the specific heat. 
In this standing point, {\it four-dimensional theories 
with inter-layer interactions are defined for every $\beta_5$  (a multi-layer world)}.

The phase diagram for $N_5 >1$ in the multi-layer world is not changed by $N_5$ 
$-\!\!\!-$ see Fig.\ref{fign5=1-8g028}.  This fact suggests us an extension 
of a multi-layer theory. 
If we try to extend this internal space to an real extra dimension, 
$N_5$ must be took to infinity. 
When $N_5$ goes to infinity, $M_k$ vanishes from (\ref{kkm}).  
To avoid it,  it is necessary for a relation among $\gamma, a_4$ and $N_5$. 
If we can introduce  a new lattice spacing, $a_5 \sim a_4 / \gamma$,    
our   squared mass (\ref{kkm}) is written as
\beq
M_k^2 \sim  \left(\frac{2\pi k}{N_5 a_5} \right)^2 \ .
\label{kkrelation}
\eeq
\noindent
This relation is  exactly  a  KK mass. Originally, our fifth direction is 
 just an internal space but (\ref{kkrelation}) shows the possibility 
that the internal space is interpreted as fifth direction. 
Although the explicit definition for $a_5$ is discussed 
in the following subsection, we note that a relation
\beq
\left(\frac{a_4}{a_5}\right)^2 \sim \frac{\beta_5}{\beta_4} = \gamma^2
\eeq
\noindent
implies just that anisotropy of gauge coupling constants corresponds to 
that  of lattice spacings like as finite temperature theories.

\subsection{Suggestion for a new definition of five-dimensional gauge theory}

In the end of the previous subsection, we saw the possibility of an extra 
dimension. To construct a continuous extra dimension, we must consider 
 the continuum limit ($a_5 \rightarrow 0$) combining with another limit 
$a_4 \rightarrow 0$. 
This is a hard parameter-tuning but it becomes  possible by the balance  
between extra-dimensional dynamics and four-dimensional one. 
In order to set the balance, we can define the quantity 
from  (\ref{Kdef}), (\ref{fourmass}) and (\ref{crit-nu}), 
\beq
{\cal R}\equiv \frac{M_k^2}{m_h^2}
= \frac{\beta_5 \lambda_2(N_5)}{\beta_4 H^2(\beta_4,\beta_5,N_5)}   
\propto \frac{\beta_5}{N_5^2 \beta_4 |\beta_5 - \beta_{5c}|^{2\nu}} \  . 
\label{defcalR}
\eeq
\noindent
If this dimensionless quantity is kept as a nonzero  finite value in the 
limit,  it is possible to make both a Higgs mass and KK one of the second 
excited mode finite.

\begin{figure}[p]
\begin{center}
\includegraphics[scale=0.5]{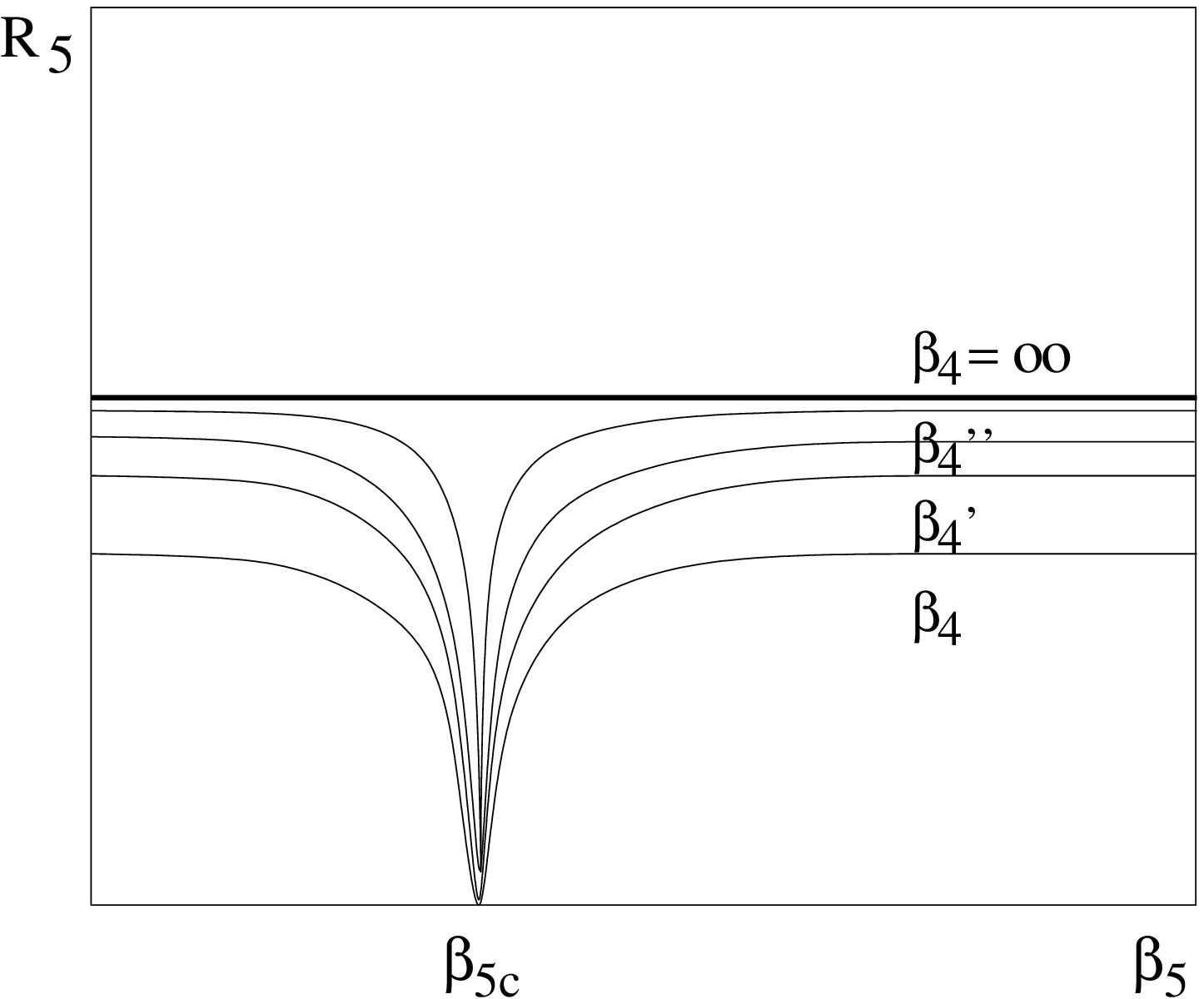}
\vspace{5cm}
\caption{ An expected envelope ($\beta_4=\infty$) formed by lines $\beta_4,\beta_4',
\beta_4''$, etc.}
\label{figR5}
\end{center}
\end{figure}

Let us consider when ${\cal R}$ can be nonzero finite.  
From the $a_4\rightarrow 0$ and 
(\ref{defcalR}), we should investigate the region 
where  $\beta_4$ is large  and $H$ is small ($\beta_5 \sim \beta_{5c}$). 
Our numerical study in section 3 suggests to us that 
the phase structure and the parameter $\beta_{5c}$ unchanged under various $N_5 >1$. 
So, we can concentrate on  large $\beta_4$ and small $H$ independent of $N_5$. 
After of all, we can introduce a scale parameter $a_5$, 
\beq
a_5 \equiv \frac{\pi}{\sqrt{\lambda_2}N_5 \gamma } a_4  . 
\eeq
\noindent
From this parameter, we can define an extra-dimensional physical scale, 

\beq
R_5 \equiv N_5 a_5  = \frac{\pi}{\sqrt{\lambda_2}\gamma} a_4   .
\eeq
Can we realize this as a five-dimensional space?  The answer shall be yes if 
the five-dimensional scale
\beq
R_5 = \sqrt{\frac{\beta_4}{\beta_5}}\frac{\pi H }{\sqrt{\lambda_2} m_h} = 
\frac{\pi}{\sqrt{{\cal R}} m_h} 
\label{rel-r5}
\eeq
\noindent
is kept nonzero finite. From the relation (\ref{rel-r5}), 
we find that $R_5$ is larger as $\beta_4$ is larger in the 
case of fixed $\beta_5$. The formed lines in Fig.\ref{figR5} 
 are expected to  make an envelope in $\beta_5 \sim \beta_{5c}$, 
which the limiting line appears in 
 the plot after $\beta_4 \rightarrow \infty$. 
{\it The existence of the envelope} shall be an important clue in the proof that 
 a five-dimensional field theory  
is constructed with a  finite extra-dimensional scale  $R_5$ 
and a finite Higgs mass. 


\section{Summary and Discussions} 

In this paper, we have investigated the phase diagram  
for a five-dimensional $SU(2)$ gauge theory. 
From the view of a layer world a la Fu-Nielsen, 
this system is  considered as an ensemble of four-dimensional gauge theories 
on layers with inter-layer interactions. 
In the region of small but nonzero $\beta_5$, the phase structure including a 
second order phase transition is surprisingly unchanged with $N_5 >1$.
In the limiting operations, we consider $a_4 \rightarrow 0$ firstly because 
we know the limit of the four-dimensional gauge theory well. 
After the construction of this multi-layer theory, 
we have sought the possibility that we can keep 
an extra-dimensional scale and a four-dimensional scale finite. 

From the universality argument, our theory on $\beta_4 \rightarrow \infty$ and 
$\beta_5=\beta_{5c}$ corresponds to a four-dimensional Ising spin model. 
The model has a second order phase transition with $\nu \geq 0.5$ \cite{V-W92,J-M-M-T-W89,B-F-M-M-P-R97,A-C-G-F-I-T-U95}.
What happens for our model in the case of  $\beta_{5c} = 0$ ? 
In $\nu > 0.5$, we can obtain a finite extra-dimensional scale
by the same procedure as above discussion.
If $\nu=0.5$, we need logarithmic corrections for non-trivial theories.  

There is another approach called deconstruction. 
It is the ensemble of asymptotic free theories  and 
has KK-like modes in four dimensions. 
Based on the construction of a four-dimensional lattice gauge theory, we  study 
the non-perturbative property such as the vacuum expectation value of 
a five-dimensional Polyakov loop.  Both approaches seem 
complimentary to each other. 

As remained problems, after detailed numerical  computations 
along to our scenario, it is natural to try (A) an extension  to matter fields,  
(B) restoration of the  rotational symmetry in the five dimensions.  
 Although, in this paper, we  have mainly investigated  
the small  $\beta_5 / \beta_4$, we may try  the connection to the large 
 $\beta_5/ \beta_4$  to solve these problems. \\

\hspace*{-0.5cm}{\Large \bf Acknowledgments}\\

\hspace*{-0.5cm}The authors thank H.~Nakano for discussions about deconstruction approaches.
We also thank S.~Ejiri and H.~Arisue for references of four-dimensional Ising models.
This work was mainly computed the Yukawa Institute Computer Facility. 
It is also supported in part by the Grants-in-Aid for Scientific 
Research No. 13135209 from the Japan Society for the Promotion of Science.

\newpage
\def\vol#1#2#3{{\bf {#1}} ({#2}) {#3}}
\def\NP{Nucl.~Phys. }
\def\PL{Phys.~Lett. }
\def\PR{Phys.~Rev. }
\def\PRL{Phys.~Rev.~Lett. }
\def\PTP{Prog.~Theor.~Phys. }
\def\MPL{Mod.~Phys.~Lett. }
\def\IJMP{Int.~J.~Mod.~Phys. }
\def\JETP{Sov.~Phys.~JETP }
\def\JP{J.~Phys. }
\def\NPPS#1#2#3{%
\NP (Proc.~Suppl. {\bf {#1}}) ({#2}) {#3}}
\def\NP{{\it Nucl}.~{\it Phys}. }
\def\PL{{\it Phys}.~{\it Lett}. }
\def\PR{{\it Phys}.~{\it Rev}. }
\def\PRL{{\it Phys}.~{\it Rev}.~{\it Lett}. }
\def\PTP{{\it Prog}.~{\it Theor}.~{\it Phys}. }
\def\MPL{{\it Mod}.~{\it Phys}.~{\it Lett}. }
\def\IJMP{{\it Int}.~{\it J}.~{\it Mod}.~{\it Phys}. }
\def\JETP{{\it Sov}.~{\it Phys}.~{\it JETP} }
\def\JP{{\it J}.~{\it Phys}. }
\def\NPPS#1#2#3{%
\NP {\it Proc}.~{\it Suppl}. {\bf {#1}} ({#2}) {#3}}
\def\th#1{hep-th/{#1}}
\def\ph#1{hep-ph/{#1}}

\clearpage

\clearpage

\clearpage

\clearpage

\clearpage

\clearpage

\clearpage

\clearpage

\clearpage

\clearpage

\clearpage

\clearpage

\clearpage

\clearpage

\clearpage

\clearpage

\clearpage

\end{document}